\documentclass[twocolumn,showpacs,amsfonts,amsmath,amssymb,floatfix,longbibliography]{revtex4-1}
\usepackage{url}
\usepackage{bm}		
\usepackage{graphicx}	
\usepackage{hyperref}

\DeclareMathOperator{\fdot}{\raisebox{.3ex}{\scalebox{1.6}{.}}}
\DeclareMathOperator{\wedgie}{\wedge}

\begin{document}


\title{A covariant Hamiltonian tetrad approach to numerical relativity}

\author{Andrew J S Hamilton}
\email{Andrew.Hamilton@colorado.edu}	
\affiliation{JILA, Box 440, U. Colorado, Boulder, CO 80309, USA}
\affiliation{Dept.\ Astrophysical \& Planetary Sciences,
U. Colorado, Boulder, CO 80309, USA}

\newcommand{\simpropto}{\raisebox{-0.7ex}[1.5ex][0ex]{$\,
                \begin{array}[b]{@{}c@{\;}} \propto \\
                [-1.5ex] \sim \end{array}$}}

\newcommand{\dd}{d}
\newcommand{\bdd}{\bm{\dd}}
\newcommand{\dext}{{\rm d}}				
\newcommand{\ddi}[1]{\dd^{#1}\mkern-1.5mu}		
\newcommand{\bddi}[1]{\bdd^{#1}\mkern-1.5mu}
\newcommand{\bddx}{\bm{\dd\mkern-2.0mu x}}
\newcommand{\DD}{D}
\newcommand{\bDD}{\bm{\DD}}
\newcommand{\Dext}{{\rm D}}				
\newcommand{\DextS}{\Dext}
\newcommand{\DextL}{\Dext_{\scriptscriptstyle\Gamma}}	
\newcommand{\LieL}{{\cal L}_{\scriptscriptstyle\Gamma}}	
\newcommand{\bcalD}{\pmb{\cal D}}
\newcommand{\ee}{e}
\newcommand{\im}{i}
\newcommand{\Ei}{{\rm Ei}}
\newcommand{\perpperp}{\perp\!\!\perp}
\newcommand{\ppartial}{\partial^2\mkern-1mu}
\newcommand{\nn}{\nonumber\\}
\newcommand{\transpose}{\top}

\newcommand{\ddual}[1]{%
  \vbox{\offinterlineskip\ialign{%
    \hfil##\hfil\cr
    $\scriptstyle\ast\mkern-2mu\ast$\cr
    \noalign{\kern.25ex}
    $#1$\cr
}}}
\newcommand{\hodge}{\mkern2mu{}^\ast\mkern-2mu}

\newcommand{\diag}{{\rm diag}}
\newcommand{\jel}{\text{\sl j}}
\newcommand{\Lz}{L}
\newcommand{\Msun}{{\rm M}_\odot}
\newcommand{\uel}{\text{\sl u}}
\newcommand{\vel}{\text{\sl v}}
\newcommand{\inn}{{\rm in}}
\newcommand{\out}{{\rm ou}}
\newcommand{\sep}{{\rm sep}}
\newcommand{\Spin}{\Sigma}

\newcommand{\Gammaop}{{\hat\Gamma}}
\newcommand{\bGammaop}{\bm{\hat\Gamma}}
\newcommand{\Rop}{{\hat R}}
\newcommand{\bRop}{\bm{\hat R}}
\newcommand{\Sop}{{\hat S}}
\newcommand{\bSop}{\bm{\hat S}}

\newcommand{\ba}{\bm{a}}
\newcommand{\bA}{\bm{A}}
\newcommand{\bb}{\bm{b}}
\newcommand{\bB}{\bm{B}}
\newcommand{\bC}{\bm{C}}
\newcommand{\be}{\bm{e}}
\newcommand{\bepsilon}{\bm{\epsilon}}
\newcommand{\bF}{\bm{F}}
\newcommand{\bg}{\bm{g}}
\newcommand{\bgamma}{\bm{\gamma}}
\newcommand{\bG}{\bm{G}}
\newcommand{\bGamma}{\bm{\Gamma}}
\newcommand{\bH}{\bm{H}}
\newcommand{\bj}{\bm{j}}
\newcommand{\bK}{\bm{K}}
\newcommand{\bmu}{\bm{\mu}}
\newcommand{\bnu}{\bm{\nu}}
\newcommand{\bp}{\bm{p}}
\newcommand{\bpi}{\bm{\pi}}
\newcommand{\bPi}{\bm{\Pi}}
\newcommand{\bq}{\bm{q}}
\newcommand{\bR}{\bm{R}}
\newcommand{\bS}{\bm{S}}
\newcommand{\bSpin}{\bm{\Spin}}
\newcommand{\bT}{\bm{T}}
\newcommand{\bvartheta}{\bm{\vartheta}}
\newcommand{\bv}{\bm{v}}
\newcommand{\bx}{\bm{x}}

\newcommand{\KCarter}{{\cal K}}
\newcommand{\Mass}{{\cal M}}
\newcommand{\mbh}{m_\bullet}
\newcommand{\Mbh}{M_\bullet}
\newcommand{\Mbhdot}{\dot{M}_\bullet}
\newcommand{\NUT}{{\cal N}}
\newcommand{\Qelec}{Q}
\newcommand{\Qelecbh}{\Qelec_\bullet}
\newcommand{\Qmag}{{\cal Q}}
\newcommand{\Qmagbh}{\Qmag_\bullet}
\newcommand{\rhosep}{\rho_{\rm Kerr}}
\newcommand{\tform}{\bar{t}}
\newcommand{\alphaform}{\bar{\alpha}}
\newcommand{\aform}{\bar{a}}
\newcommand{\unit}[1]{\, {\rm{#1}}}
\newcommand{\xin}{x_{\rm in}}
\newcommand{\zeroform}{\bar{0}}

\newcommand{\elec}{{\rm e}}
\newcommand{\grav}{{\rm g}}
\newcommand{\mat}{{\rm m}}

\hyphenpenalty=3000

\newcommand{\ntab}{
    \begin{table}
    \caption[]{
Numbers of equations and variables
    }
    \label{ntab}
    \begin{ruledtabular}
    \begin{tabular}{lcccc}
Approach & Eqs.\ mot. & Constraints & Identities & Variables \\
This paper & 24 & 10 & 6 & 40 \\
ADM & 12 & 10 & 18 & 40 \\
BSSN & 15 & 10 & 15 & 40 \\
WEBB & 30 & 16 & 0 & 46 \\
LQG & 18 & 7 & 0 & 25
    \end{tabular}
    \end{ruledtabular}
    \end{table}
}

\begin{abstract}
A Hamiltonian approach to the equations of general relativity is proposed
using the powerful mathematical language of
multivector-valued differential forms.
In the approach, the gravitational coordinates are the
12 spatial components of the line interval (the vierbein)
including their antisymmetric parts,
and their 12 conjugate momenta.
A feature of the proposed formalism is that it allows Lorentz
gauge freedoms to be imposed on the Lorentz connections rather than
on the vierbein, which may facilitate numerical integration in
some challenging problems.
The 40 Hamilton's equations comprise
$12 + 12 = 24$ equations of motion,
10 constraint equations (first class constraints,
which must be arranged on the initial hypersurface of constant time,
but which are guaranteed thereafter by conservation laws),
and 6 identities (second class constraints).
The 6 identities define a trace-free spatial tensor
that is the gravitational analog of the magnetic field of electromagnetism.
If the gravitational magnetic field is promoted to an independent field
satisfying its own equation of motion,
then the system becomes the WEBB system,
which is known to be strongly hyperbolic.
Some other approaches,
including ADM, BSSN, WEBB, and Loop Quantum Gravity,
are translated into the language of multivector-valued forms,
bringing out their underlying mathematical structure.
\end{abstract}


\date{\today}

\maketitle

\section{Introduction}


The motivation for this paper is the problem of the interior structure
of astronomically realistic rotating black holes.
The inner horizons of rotating black holes are subject to the inflationary
instability discovered by Poisson \& Israel
\cite{Poisson:1989zz}.
The inflationary instability is the nonlinear consequence of the
infinite blueshift at the inner horizon first pointed out by Penrose
\cite{Penrose:1968}.
Numerical computations presented in a subsequent paper
\cite{Hamilton:2017qls}
indicate that in a rotating black hole that accretes, as real black holes do,
inflation stalls, and Belinskii-Khalatnikov-Lifshitz (BKL) collapse
\cite{Belinskii:1982}
to a spacelike singular surface ensues.

Calculating the interior structure of realistic rotating black holes
presents severe numerical challenges.
One challenge is that gradients of the metric grow huge.
Another is that collapse is intrinsically unstable to the growth
of anisotropies.
The numerical method must be robust enough to follow physical
instability without gauge instability.

A basic idea behind the present paper is that the physical environment
may provide a preferred set of Lorentz frames,
which in turn define a preferred set of observers.
For example, in the world familiar to humans,
the Cosmic Microwave Background (CMB) defines preferred frames,
and selects preferred (comoving, not freely-falling) observers
who remain at rest with respect to the CMB.
Similarly, in the interior of a rotating black hole,
accreting streams focus along each of the outgoing and ingoing
principle null directions as they approach the inner horizon.
The null directions along which the accreting streams focus
define preferred frames analogous to the CMB.

Selecting the preferred frames may involve imposing gauge conditions
on the Lorentz connections
(equivalently, on derivatives of the line interval).
The approach proposed in the present paper is designed to allow
a broader range of freedoms in choosing gauge conditions,
including the freedom to impose gauge freedoms
on the Lorentz connections as well as the line interval.
For example,
singularity theorems
\cite{Senovilla:1997}
are built on geodesic congruences
in which geodesic gauge conditions are imposed on the Lorentz connections.

General relativity is a gauge theory of two local symmetries
\cite{Nester:2016rsy,Blagojevic:2013},
the 4-dimensional group of translations
(coordinate transformations),
and the 6-dimensional group of Lorentz transformations.
Most treatments of numerical relativity work only with
Lorentz gauge-invariant quantities,
namely the metric and its first and second coordinate derivatives.
The present paper follows the alternative strategy of keeping
both coordinate and Lorentz symmetries distinct and manifest,
the tetrad approach.
Tetrad-based approaches to numerical relativity
are reviewed by \cite{Bardeen:2011ip}.
Because of the complexity of the symmetry group,
the gravitational equations
written out in index notation
tend to degenerate into a mess
(what Cartan \cite[Preface]{Cartan:1951}
called the ``d\'ebauches d'indices''),
that obscures their underlying structure.
To keep the structure transparent,
it is highly advantageous to use the abstract mathematical language of
multivector-valued differential forms.
A multivector is an element of the Clifford algebra (geometric algebra)
generated by vectors
$\bgamma_m \equiv \{ \bgamma_0 , \bgamma_1 , \bgamma_2 , \bgamma_3 \}$
with inner product the Minkowski metric,
$\bgamma_m \cdot \bgamma_n = \eta_{mn}$,
and an antisymmetric outer product (wedge product).
The reader is assumed to be familiar with both multivectors
(see \cite{Doran:2003})
and forms
\cite{Cartan:2001,Ivey:2016}.
In this paper,
local Lorentz transformations transform multivectors
while keeping forms unchanged;
while coordinate transformations transform forms
while keeping multivectors unchanged.

Following Dirac \cite{Dirac:1964},
the system of Hamilton's equations for a constrained Hamiltonian system,
which includes local gauge theories such as general relativity,
divides into three groups:
(1) equations of motion,
which are evolution equations that propagate the coordinates
and their conjugate momenta forward in time;
(2) constraint equations
(first class constraints, in Dirac's terminology),
which must be arranged to be satisfied
on the initial hypersurface of constant time,
but which are guaranteed thereafter by conservation laws;
and
(3) identities
(second class constraints, in Dirac's terminology),
which define redundant fields in terms of others.
The constraints and identities share the property that they
involve only spatial, not time, derivatives of the coordinates and momenta.

The number of constraint equations
equals the dimensionality of the symmetry group of the gauge theory,
which in the case of general relativity
is 10, the dimensionality of the Poincar\'e group.
The constraint equations of general relativity are commonly called
Hamiltonian and momentum constraints,
which arise from symmetry under local translations (coordinate transformations),
and Gaussian constraints,
which arise from symmetry under local Lorentz transformations.

In general relativity,
there are altogether 40 gravitational Hamilton's equations,
including the 10 constraint equations.
In the approach proposed in this paper,
the remaining 30 Hamilton's equations divide into
24 equations of motion and 6 identities.
The 6 identities define a trace-free spatial tensor
that is the gravitational equivalent of the magnetic field of electromagnetism.
If the gravitational magnetic field is promoted to an independent field
satisfying its own equation of motion,
then the system becomes the
Wahlquist-Estabrook-Buchman-Bardeen
(WEBB)
\cite{Estabrook:1996wa,Buchman:2003sq,Buchman:2005ub}
system.

The version of general relativity considered in this paper is
actually Einstein-Cartan theory,
which is the extension of general relativity to include torsion.
Standard general relativity is recovered by setting torsion to zero;
but from a fundamental perspective setting torsion to zero is,
as Hehl \cite{Hehl:2012pi,Blagojevic:2013} has emphasized, artificial.
Spin-$\tfrac{1}{2}$ fields generate torsion:
fermions have spin angular-momentum that sources nonzero torsion.
Torsion,
whose effects are negligible under normal (accessible to humans) conditions,
could potentially play a role under the extreme conditions attending inflation
and collapse inside black holes.

The approach is presented in \S\ref{grav-sec},
and its integrability assessed in \S\ref{hyperbolicity-sec}.
The approach is compared to a number of other approaches,
including ADM, BSSN, WEBB, and Loop Quantum Gravity,
in \S\ref{admbssnlqg-sec}.
To keep the main text compact,
some details of the notation (\S\ref{notation-sec}),
along with some extra material
on the Bianchi identities and conservation laws (\S\ref{bianchi-sec}),
on Cartan's Test (\S\ref{cartantest-sec}),
and on electromagnetism (\S\ref{electromagnetism-sec}),
are deferred to Appendices.

\section{Gravitational equations}
\label{grav-sec}

The structure of the gravitational equations derived in this section
is quite similar to the structure of the equations of electromagnetism,
except that the gravitational equations are more complicated.
In particular,
the decomposition of Hamilton's equations into
equations of motion, constraints, and identities
in the two theories is quite similar.
For reference, Appendix~\ref{electromagnetism-sec}
presents an exposition of electromagnetism
that brings out the similarity of structure.

\subsection{Line interval, connection, torsion, and curvature}

The degrees of freedom of the gravitational field are contained in
the line interval $\be$,
or vierbein,
a vector 1-form with $4 \times 4 = 16$ degrees of freedom,
and
the Lorentz connection $\bGamma$,
a bivector 1-form with $6 \times 4 = 24$ degrees of freedom,
\begin{subequations}
\label{eGammaformdef}
\begin{align}
\label{eformdef}
  \be
  &\equiv
  e_{k{\kappa}}
  \, \bgamma^k
  \, \dd x^{\kappa}
  \ ,
\\
\label{Gammaformdef}
  \bGamma
  &\equiv
  \Gamma_{kl{\kappa}}
  \,
  \bgamma^k \wedgie \bgamma^l
  \,
  \dd x^{\kappa}
  \ ,
\end{align}
\end{subequations}
implicitly summed over distinct antisymmetric sequences of paired indices.
The components $\Gamma_{kl\kappa}$ of the Lorentz connections,
defined by the rate of change of the tetrad vectors $\bgamma_l$ along directions
$\dd x^\kappa$ relative to parallel transport,
\begin{equation}
  \Gamma_{kl\kappa}
  \equiv
  \bgamma_k \cdot {\partial \bgamma_l \over \partial x^\kappa}
  \ ,
\end{equation}
are sometimes called Ricci rotation coefficients,
or (especially when expressed in a Newman-Penrose double-null tetrad basis)
spin coefficients.
Whereas the vierbein coefficients $e_{k\kappa}$ are tensors with respect
to both coordinate and Lorentz transformations,
the Lorentz connection coefficients $\Gamma_{kl\kappa}$
are coordinate tensors but not Lorentz tensors.

It is convenient to use the notation $\be^p$
to denote the normalized $p$-volume element
\begin{equation}
\label{dpx}
  \be^p
  \equiv
  {1 \over p!} \,
  \overbrace{\be \wedgie \cdots \wedgie \be}^{\mbox{$p$ terms}}
  \ ,
\end{equation}
which is both a $p$-form and a grade-$p$ multivector.
The factor $1/p!$ compensates for multiple counting of
distinct indices, so that $\be^p$ has the correct normalization
for a $p$-volume element.
For example, $\be^2$ is the 2-volume element, or area element,
while $\be$ is the 1-volume element, or line interval.
The scalar product of a multivector
$\ba \equiv a_K \, \bgamma^K$
of grade $p$ with the $p$-volume element $\be^p$ is the $p$-form
$a_\Lambda \, \ddi{p} x^\Lambda$,
\begin{equation}
  \ba \cdot \be^p
  =
  a_\Lambda \, \ddi{p} x^\Lambda
  \ ,
\end{equation}
implicitly summed over distinct antisymmetric sequences $\Lambda$
of $p$ coordinate indices.

The commutator of the (coordinate and Lorentz) covariant derivative defines
the torsion vector 2-form $\bS$
and the Riemann curvature bivector 2-form $\bR$
in terms of exterior derivatives of the line interval and connection,
\begin{subequations}
\label{Cartan}
\begin{align}
\label{SCartan}
  \bS
  &\equiv
  \dext \be
  +
  \tfrac{1}{2}
  [ \bGamma , \be ]
  \ ,
\\
\label{RCartan}
  \bR
  &\equiv
  \dext \bGamma + \tfrac{1}{4} [ \bGamma , \bGamma ]
  \ ,
\end{align}
\end{subequations}
which are Cartan's equations of structure.
Cartan's equations~(\ref{Cartan})
are definitions of torsion and curvature,
not equations of motion.
Equations of motion follow from varying the action.

The exterior derivative $\dext$ is coordinate covariant
but not Lorentz covariant.
The (coordinate and Lorentz) covariant exterior derivative
of any multivector form $\ba$ is
\begin{equation}
  \Dext \ba
  =
  \dext \ba
  +
  \tfrac{1}{2}
  [ \bGamma , \ba ]
  \ ,
\end{equation}
the last term accounting for the change in the tetrad vectors
relative to parallel transport.
Thus the right hand side of the definition~(\ref{SCartan}) of torsion
is the covariant exterior derivative $\Dext \be$ of the line interval,
which is a (coordinate and Lorentz) tensor.
The right hand side of the definition~(\ref{RCartan}) of curvature does not
have the appearance of a covariant exterior derivative,
but $\bGamma$ is not a Lorentz tensor,
and the definition~(\ref{RCartan}) is precisely such as to make
the Riemann curvature $\bR$ a (coordinate and Lorentz) tensor.

\subsection{Gravitational action}

The Hilbert action is
\begin{equation}
\label{Sgform}
  S_\grav
  \equiv
  \int L_\grav
  \ ,
\end{equation}
where $L_\grav$ is the Hilbert Lagrangian scalar 4-form
\begin{equation}
\label{Lgform}
  L_\grav
  \equiv
  -
  {I \over \kappa} \,
  \be^2
  \wedgie
  \bR
  =
  -
  {I \over \kappa} \,
  \be^2
  \wedgie
  \left(
  \dext \bGamma + \tfrac{1}{4} [ \bGamma , \bGamma ]
  \right)
  \ ,
\end{equation}
where $I$ is the pseudoscalar, equation~(\ref{pseudoscalar}),
and $\kappa \equiv 8\pi G$ is the normalized Newton's gravitational constant.

The Hilbert Lagrangian~(\ref{Lgform}) is in super-Hamiltonian
(in the terminology of \cite{MTW:1973})
form
$I \bp \wedgie \dext \bq - H_\grav$
with gravitational coordinates the Lorentz connection $\bq = \bGamma$,
conjugate momenta the area element $\bp = - \be^2 / \kappa$,
and super-Hamiltonian
\begin{equation}
\label{Hgform}
  H_\grav
  =
  {I \over \kappa} \, \be^2 \wedgie \tfrac{1}{4} [ \bGamma , \bGamma ]
  \ .
\end{equation}
However, the area element $\be^2$,
a bivector 2-form with $6 \times 6 = 36$ components,
has excess degrees of freedom compared to
the 16-component line interval $\be$.
A (standard) solution to the excess degrees of freedom
is to integrate the Hilbert action by parts
so that the gravitational variables become the line interval $\be$
and its conjugate momentum $\bpi$,
the pseudovector 2-form defined by
\begin{equation}
\label{piformdef}
  \bpi
  \equiv
  -
  \be \wedgie \bGamma
  \ .
\end{equation}
Specifically, the term $- \be^2 \wedgie \dext \bGamma$
in the Hilbert Lagrangian~(\ref{Lgform})
integrates by parts to
\begin{equation}
  - \be^2 \wedgie \dext \bGamma
  =
  \dext \bvartheta
  +
  \dext \be \wedgie ( \be \wedgie \bGamma )
  \ .
\end{equation}
where $\bvartheta$ is the expansion
defined by equation~(\ref{expansionform}).
The total exterior derivative term $\dext \bvartheta$ is the
Gibbons-Hawking-York
\cite{York:1972sj,Gibbons:1976ue}
boundary term.
With the boundary term discarded,
the integration by parts brings the Hilbert Lagrangian~(\ref{Lgform}) to
\begin{align}
\label{Lgaltform}
  L_\grav^\prime
  =
  L_\grav - \dext \bvartheta
  &=
  -
  {I \over \kappa}
  \left(
  \be \wedgie \bGamma
  \wedgie
  \dext \be
  +
  \tfrac{1}{4}
  \be^2
  \wedgie
  [ \bGamma , \bGamma ]
  \right)
\nonumber
\\
  &=
  {I \over \kappa} \,
  \bpi \wedgie
  \left(
  \dext \be
  +
  \tfrac{1}{4}
  [ \bGamma , \be ]
  \right)
  \ .
\end{align}
The modified Hilbert Lagrangian~(\ref{Lgaltform}) is in super-Hamiltonian form
with gravitational coordinates the line interval $\be$,
conjugate momenta $\bpi / \kappa$,
and the same super-Hamiltonian~(\ref{Hgform}) as before.

The Lorentz connection $\bGamma$ and conjugate momentum $\bpi$
are invertibly related,
equation~(\ref{piformdef}) inverting to
\begin{equation}
\label{Gammapi}
  \bGamma
  =
  - \,
  \ddual\bpi{}^\transpose
  +
  \be \wedgie
  \ddual{\bvartheta}{}^\transpose
  \ ,
\end{equation}
where $\ddual{\phantom{a}}$ denotes the double dual, equation~(\ref{ddualdef}),
${}^\transpose$ the transpose operation~(\ref{formtranspose}),
and $\bvartheta$ is the expansion defined by equation~(\ref{expansionform}).

Variation of the gravitational action $S_\grav^\prime$
with the modified Hilbert Lagrangian~(\ref{Lgaltform}) yields
\begin{equation}
\label{dSgaltformSR}
  \delta S_\grav^\prime
  =
  {I \over \kappa}
  \oint
  \bpi
  \wedgie
  \delta \be
  +
  {I \over \kappa}
  \int
  \delta \bpi
  \wedgie
  \bS
  -
  \bPi
  \wedgie
  \delta \be
  \ ,
\end{equation}
where the curvature pseudovector 3-form $\bPi$ is defined to be
\begin{equation}
\label{Piformdef}
  \bPi
  \equiv
  \be \wedgie \bR
  -
  \bS \wedgie \bGamma
  =
  \dext \bpi
  +
  \tfrac{1}{2} [ \bGamma , \bpi ]
  -
  \tfrac{1}{4}
  \be \wedgie
  [ \bGamma , \bGamma ]
  \ .
\end{equation}
The $\be \wedgie \bR$ term in $\bPi$ equals the double dual
of the vector 1-form Einstein tensor $\bG$,
\begin{equation}
  \ddual{\bG}
  =
  \be \wedgie \bR		
  \ .
\end{equation}

For any given species of matter,
the spin angular-momentum $\bSpin$, a bivector 1-form,
and the matter energy-momentum $\bT$, a vector 1-form,
are conventionally defined by the variation of the matter action
with respect to respectively the connection $\bGamma$
and the line interval $\be$
\cite{Blagojevic:2013}.
In multivector-forms language,
\begin{equation}
\label{dSmform}
  \delta S_\mat
  =
  I
  \int
  \ddual{\bSpin} \wedgie \delta \bGamma
  +
  \delta \be \wedgie \ddual{\bT}
  \ ,
\end{equation}
where $\ddual{\bSpin}$ and $\ddual{\bT}$
are the double duals
of the spin angular-momentum
$\bSpin$
and energy-momentum
$\bT$
of the matter.
The spin angular-momentum is so called because it vanishes for gauge fields
such as electromagnetism,
but is finite for spinor fields
\cite{Blagojevic:2013}.

Variation of the matter action $S_\mat$ instead with respect to
$\delta \bpi$ and $\delta \be$
defines modified versions
$\tilde{\bSpin}$
and
$\tilde{\bT}$
of the spin angular-momentum and the energy-momentum,
\begin{equation}
\label{dSmaltform}
  \delta S_\mat
  =
  I
  \int
  - \,
  \delta \bpi
  \wedgie
  \tilde{\bSpin}
  +
  \tilde{\bT}
  \wedgie
  \delta \be
  \ .
\end{equation}
The original and modified
spin angular-momentum and energy-momenta of matter are invertibly related by
\begin{subequations}
\label{torsionenergymomentummodified}
\begin{align}
\label{torsionmodified}
  \ddual{\bSpin}
  &=
  \be \wedgie \tilde{\bSpin}
  \ ,
\\
\label{energymomentummodified}
  \ddual{\bT}
  &=
  \tilde{\bT}
  -
  \tilde{\bSpin}
  \wedgie
  \bGamma
  \ .
\end{align}
\end{subequations}

The combined variation of gravitational and matter actions yields
Hamilton's equations for the torsion $\bS$ and curvature $\bPi$,
\begin{subequations}
\label{eqaltform}
\begin{align}
\label{torsioneqaltform}
  \bS
  &=
  \kappa
  \tilde{\bSpin}
  \ ,
\\
\label{Einsteineqaltform}
  \bPi
  &=
  \kappa
  \tilde{\bT}
  \ .
\end{align}
\end{subequations}
Equation~(\ref{Einsteineqaltform}) is Einstein's equation (with torsion).
More explicitly,
Hamilton's equations~(\ref{eqaltform}) are
\begin{subequations}
\label{eqaltformd}
\begin{align}
\label{torsioneqaltformd}
  \dext \be + \tfrac{1}{2} [ \bGamma , \be ]
  &=
  \kappa \tilde{\bSpin}
  \ ,
\\
\label{Einsteineqaltformd}
  \dext \bpi + \tfrac{1}{2} [ \bGamma , \bpi ]
  -
  \tfrac{1}{4}
  \be \wedgie [ \bGamma , \bGamma ]
  &=
  \kappa \tilde{\bT}
  \ .
\end{align}
\end{subequations}

In most problems in general relativity, the spin angular-momentum
is taken to vanish,
$\bSpin = 0$,
and thus torsion vanishes,
$\bS = 0$,
as indeed standard general relativity assumes.
Spin angular-momentum is retained here however for completeness.

\subsection{Expansion}

The contraction of the momentum $\bpi$ defines the
expansion $\bvartheta$,
a pseudoscalar 3-form with 4 components,
\begin{equation}
\label{expansionform}
  \bvartheta
  \equiv
  \tfrac{1}{2} \be \wedgie \bpi
  =
  -
  \be^2 \wedgie \bGamma
  \ .
\end{equation}
The transpose of the double dual of the expansion,
which enters equation~(\ref{Gammapi}), is
\begin{equation}
  \ddual{\bvartheta}{}^\transpose
  =
  \Gamma^p_{kp} \, \bgamma^k
  \ .
\end{equation}
After a 3{+}1 space-time split, \S\ref{split-sec},
the tetrad time component
$\Gamma^p_{0p}$
is what is commonly called the trace of the extrinsic curvature,
or three times the expansion,
justifying the nomenclature.

The exterior derivative of the 3-volume $\be^3$
is the pseudovector 4-form $\dext \be^3$ satisfying,
on substitution of Hamilton's equation~(\ref{torsioneqaltform}),
\begin{equation}
\label{dvolume}
  \dext \be^3 + \tfrac{1}{2} [ \bGamma , \be^3 ]
  =
  \dext \be^3 - \be \cdot \bvartheta
  =
  \kappa
  \be^2 \wedgie \tilde{\bSpin}
  \ .
\end{equation}
The term $\be \cdot \bvartheta$ equals the transpose of the expansion,
\begin{equation}
  \be \cdot \bvartheta
  =
  \bvartheta^\transpose
  \ .
\end{equation}

The exterior derivative of the expansion $\bvartheta$
is the pseudoscalar 4-form $\dext \bvartheta$ satisfying,
on substitution of Hamilton's equations~(\ref{eqaltformd}),
\begin{align}
\label{dexpansionform}
  \dext \bvartheta
  +
  \tfrac{1}{4}
  \be^2
  \wedgie
  [ \bGamma , \bGamma ]
  &=
  -
  {\kappa \over 2}
  \bigl(
  \be \wedgie \tilde{\bT}
  +
  \tilde{\bSpin} \wedgie \bpi
  \bigr)
\\
\nonumber
  &=
  -
  \kappa
  \bigl(
  \tfrac{1}{2}
  \ddual{T}
  +
  \tilde{\bSpin}
  \wedgie \bpi
  \bigr)
  \ ,
\end{align}
where $\ddual{T} \equiv \be \wedgie \ddual{\bT}$
is the double dual of the trace $T$ of the energy-momentum tensor $\bT$.

\subsection{3+1 split}
\label{split-sec}

The torsion equation~(\ref{torsioneqaltformd}) comprises, apparently,
24 equations for the 16 components of the line interval $\be$,
while the Einstein equation~(\ref{Einsteineqaltformd}) comprises
16 equations for the 24 components of the conjugate momentum $\bpi$.
The apparent mismatch between the number of equations and degrees of freedom
is not a practical barrier to solving Hamilton's equations~(\ref{eqaltformd}).
The 24 torsion equations can be reinterpreted as defining the
24 components of the connection $\bGamma$
(and hence $\bpi$, equation~(\ref{piformdef}))
in terms of the spin angular-momentum $\tilde{\bSpin}$
and the exterior derivative of the line interval $\be$;
and then the 16 Einstein equations are second-order differential equations
for the line interval.
This is the Lagrangian (second-order) approach.

However,
it is advantageous to pursue the Hamiltonian (first-order) approach,
since it opens the door to the powerful apparatus of canonical transformations,
sheds light on the fundamental structure of the gravitational equations,
and offers a possible pathway to quantization.

The apparent mismatch between equations and degrees of freedom stems
from two reasons.
First,
the 16-component line interval $\be$ and 24-component momentum $\bpi$
have excess degrees of freedom arising from the 4 coordinate and 6
Lorentz gauge degrees of freedom.
Second,
a covariant description requires redundant fields
that can be expressed entirely in terms of other fields.
The simplest example of a redundant field is the magnetic field
of electromagnetism.
In electromagnetism, after a 3+1 split
(see \S\ref{splitelecform-sec}),
the coordinates are the 3 spatial components $\bA$
of the electromagnetic potential 1-form,
and their conjugate momenta are the electric field,
the 3 spatial components $\hodge{\bF}$
of the dual electromagnetic field.
The remaining 3 components of the electromagnetic field,
comprising the magnetic field,
the 3 spatial components $\bF$,
are redundant because they equal the spatial exterior derivative $\dext \bA$
of the spatial components of the potential;
but the magnetic field nevertheless constitutes part
of the 6-component 4-dimensional covariant electromagnetic field.

The standard approach to isolating excess degrees of freedom is,
following Dirac \cite{Dirac:1964},
to perform a 3+1 space+time split of the equations,
and to interpret only those equations involving time derivatives
as genuine equations of motion.
The remaining equations are either constraint equations
(first class constraints),
which are equations that must be arranged to be satisfied in the initial
conditions but which are thereafter guaranteed by conservation laws,
or else identities
(second class constraints),
which define redundant fields in terms of others.

In splitting a multivector form $\ba$ into time and space components,
it is convenient to adopt a notation in which
$\ba_{\tform}$
(subscripted ${\tform}$)
represents all the coordinate time $t$ parts of the form,
while
$\ba_{\alphaform}$
(subscripted ${\alphaform}$)
represents the remaining spatial coordinate components.
The $\tform$ and $\alphaform$
subscripts should be interpreted as labels, not indices.
Thus a multivector $p$-form $\ba$ splits as
\begin{equation}
\label{formsplit}
  \ba =
  \ba_{\tform} + \ba_{\alphaform}
  \equiv
  a_{K{tA}} \, \bgamma^K
  \, \ddi{p} x^{tA}
  +
  a_{K{B}} \, \bgamma^K
  \, \ddi{p} x^{B}
  \ ,
\end{equation}
implicitly summed over distinct antisymmetric sequences
$K$ of Lorentz indices and $A$ and $B$ of spatial coordinate indices.
Note that
only the coordinate components are split:
the Lorentz components are {\em not\/} split into time and space parts.
The time component of a product
(geometric product of multivectors, exterior product of forms)
of any two multivector forms $\ba$ and $\bb$ satisfies
\begin{equation}
  ( \ba \bb )_{\tform}
  =
  \ba_{\tform} \bb_{\alphaform}
  +
  \ba_{\alphaform} \bb_{\tform}
\end{equation}
with no minus signs
(minus signs from the antisymmetry of form indices
cancel minus signs from commuting $\dd t$ through a spatial form).

Splitting the variation~(\ref{dSgaltformSR})
of the gravitational action
into time and space parts gives
\begin{align}
\label{dSgaltformtSR}
  \delta S_\grav^\prime
  &=
  {I \over \kappa}
  \oint_{t_{\rm i}}^{t_{\rm f}}
  ( \bpi \wedgie \delta \be )_{\tform}
  +
  {I \over \kappa}
  \left[
  \oint
  ( \bpi \wedgie \delta \be )_{\alphaform}
  \right]_{t_{\rm i}}^{t_{\rm f}}
\\
\nonumber
  &+
  {I \over \kappa}
  \int
  \delta \bpi_{\alphaform}
  \wedgie
  \bS_{\tform}
  +
  \delta \bpi_{\tform}
  \wedgie
  \bS_{\alphaform}
  -
  \bPi_{\alphaform}
  \wedgie
  \delta \be_{\tform}
  -
  \bPi_{\tform}
  \wedgie
  \delta \be_{\alphaform}
  \ .
\end{align}
After the 3+1 split,
the gravitational coordinates are the 12 spatial components $\be_{\alphaform}$
of the line interval,
and their conjugate momenta are the 12 spatial components $\bpi_{\alphaform}$.
The two surface integrals in equation~(\ref{dSgaltformtSR})
are first an integral over time at the spatial boundary,
and second a difference of integrals over spatial caps
at the initial and final times $t_{\rm i}$ and $t_{\rm f}$.

Variation of the combined gravitational and matter actions
with respect to the variations
$\delta \be_{\alphaform}$ and $\delta \bpi_{\alphaform}$
of the spatial coordinates and momenta yields $12 + 12 = 24$ equations of motion
involving time derivatives,
\begin{subequations}
\label{eqaltformtRS}
\begin{alignat}{2}
\label{torsionaltformtS}
  \mbox{12 eqs of mot:}
  &\quad&
  \bS_{\tform}
  &=
  \kappa \tilde{\bSpin}_{\tform}
  \ ,
\\
\label{EinsteineqaltformtR}
  \mbox{12 eqs of mot:}
  &\quad&
  \bPi_{\tform}
  &=
  \kappa \tilde{\bT}_{\tform}
  \ .
\end{alignat}
\end{subequations}
Equations~(\ref{eqaltformtRS}) are equations of motion in the sense
that they determine the time derivatives
$\dext_{\tform} \be_{\alphaform}$
and
$\dext_{\tform} \bpi_{\alphaform}$
of the gravitational coordinates and momenta.
The time derivative here is the 1-form
$\dext_{\tform} \equiv ( \partial / \partial t ) \dd t$.
More explicitly, the equations of motion~(\ref{eqaltformtRS}) are
\begin{subequations}
\label{eqaltformt}
\begin{align}
\label{torsioneqaltformt}
  \dext_{\tform} \be_{\alphaform} + \tfrac{1}{2} [ \bGamma_{\tform} , \be_{\alphaform} ]
  + \dext_{\alphaform} \be_{\tform} + \tfrac{1}{2} [ \bGamma_{\alphaform} , \be_{\tform} ]
  &=
  \kappa \tilde{\bSpin}_{\tform}
  \ ,
\\
\label{Einsteineqaltformt}
  \dext_{\tform} \bpi_{\alphaform} + \tfrac{1}{2} [ \bGamma_{\tform} , \bpi_{\alphaform} ]
  + \dext_{\alphaform} \bpi_{\tform} + \tfrac{1}{2} [ \bGamma_{\alphaform} , \bpi_{\tform} ]
  &
  -
  \tfrac{1}{4}
  \left( \be \wedgie [ \bGamma , \bGamma ] \right)_{\tform}
\nonumber
\\
  &=
  \kappa \tilde{\bT}_{\tform}
  \ .
\end{align}
\end{subequations}

Variation of the action with respect to the variations
$\delta \be_{\tform}$ and $\delta \bpi_{\tform}$
of the time components of the coordinates and momenta yields equations
involving only spatial coordinate derivatives, no coordinate time derivatives.
These purely spatial equations are not equations of motion;
rather they are either constraints,
if their continued satisfaction is guaranteed by a conservation law,
or else identities.

Variation of the action with respect to the 4-component time component
$\delta \be_{\tform}$ of the line interval
yields the 4 Hamiltonian/momentum constraints,
\begin{equation}
\label{EinsteinconstraintfaltformtR}
  \mbox{4 Ham/mom:}
  \quad
  \bPi_{\alphaform}
  =
  \kappa \tilde{\bT}_{\alphaform}
  \ ,
\end{equation}
or more explicitly
\begin{equation}
\label{Einsteinconstraintfaltformt}
  \mbox{4 Ham/mom:}
  \quad
  \dext_{\alphaform} \bpi_{\alphaform} + \tfrac{1}{2} [ \bGamma_{\alphaform} , \bpi_{\alphaform} ]
  -
  \tfrac{1}{4}
  \left( \be \wedgie [ \bGamma , \bGamma ] \right)_{\alphaform}
  =
  \kappa \tilde{\bT}_{\alphaform}
  \ .
\end{equation}
The 4 components of $\be_{\tform}$,
which are commonly called (minus) the lapse and shift,
equation~(\ref{lapseshift}),
are arbitrarily adjustable by a coordinate transformation.
Associated with this symmetry is a conservation law of energy-momentum
(Appendix~\ref{conservationenergyomomentum-sec}).
The conservation law guarantees that if the
Hamiltonian/momentum constraints~(\ref{EinsteinconstraintfaltformtR})
are arranged to be satisfied on the initial hypersurface of constant time $t$,
then they will satisfied thereafter.
It should be noted that the mere fact that $\be_{\tform}$ {\em can\/}
be treated as a gauge variable
arbitrarily adjustable by a coordinate transformation
does not mean that $\be_{\tform}$ {\em must\/}
be treated as a gauge variable.
But regardless of the choice of gauge, there are always 4 constraints
whose ongoing validity is guaranteed by conservation of energy-momentum.

Variation of the action with respect to the 12-component
time component
$\delta \bpi_{\tform}$
of the conjugate momentum yields 6 Gaussian constraints
(first class constraints)
and 6 identities
(second class constraints),
\begin{equation}
\label{torsionconstraintfaltformtS}
  \mbox{6 Gauss + 6 ids:}
  \quad
  \bS_{\alphaform}
  =
  \kappa \tilde{\bSpin}_{\alphaform}
  \ ,
\end{equation}
or more explicitly
\begin{equation}
\label{torsionconstraintfaltformt}
  \mbox{6 Gauss + 6 ids:}
  \quad
  \dext_{\alphaform} \be_{\alphaform} + \tfrac{1}{2} [ \bGamma_{\alphaform} , \be_{\alphaform} ]
  =
  \kappa \tilde{\bSpin}_{\alphaform}
  \ ,
\end{equation}

The 6 Gaussian constraints
in equations~(\ref{torsionconstraintfaltformt})
are associated with symmetry under the
6 degrees of freedom of Lorentz transformations.
There is some freedom in choosing gauge variables
arbitrarily adjustable under a Lorentz transformation.
One choice is to treat the
6-component time component $\bGamma_{\tform}$
of the Lorentz connection bivector 1-form
as an arbitrarily adjustable gauge variable.
The Gaussian constraints then follow from varying the action with respect
to $\delta \bGamma_{\tform}$,
and comprise the subset of
equations~(\ref{torsionconstraintfaltformtS})
obtained by contracting with the spatial line interval $\be$,
\begin{equation}
\label{GaussianconstraintfaltformtS}
  \mbox{6 Gauss:}
  \quad
  ( \be \wedgie \bS )_{\alphaform}
  =
  \kappa ( \be \wedgie \tilde{\bSpin} )_{\alphaform}
  \ .
\end{equation}
Associated with the Lorentz symmetry is a law of conservation of
angular momentum
(Appendix~\ref{conservationangularmomentum-sec}).
This conservation law guarantees that if
the Gaussian constraints~(\ref{GaussianconstraintfaltformtS})
are arranged to be satisfied on the initial hypersurface of constant time,
then the constraints will continue to be satisfied thereafter.
The Gaussian constraints~(\ref{GaussianconstraintfaltformtS})
may be rewritten
\begin{equation}
\label{GaussianconstraintpaltformtS}
  \mbox{6 Gauss:}
  \quad
  \dext_{\alphaform} \be^2_{\alphaform}
  -
  \be_{\alphaform} \cdot \bpi_{\alphaform}
  =
  \kappa ( \be \wedgie \tilde{\bSpin} )_{\alphaform}
  \ ,
\end{equation}
which is a differential relation between the spatial coordinates and momenta
$\be_{\alphaform}$ and $\bpi_{\alphaform}$,
whose evolution is determined by their equations of motion~(\ref{eqaltformt}).

The 6 identities
in equations~(\ref{torsionconstraintfaltformt})
comprise its trace-free spatial part.
Projected into the coordinate frame,
the 9 spatial components
$S_{\gamma\alpha\beta}$
of the torsion are
\begin{equation}
  S_{\gamma\alpha\beta}
  =
  e^k{}_\gamma
  S_{k\alpha\beta}
  \ .
\end{equation}
The 3-component spatial trace of the spatial coordinate-frame torsion is
\begin{equation}
  S^\beta_{\alpha\beta}
  \equiv
  g^{\beta\gamma}
  S_{\gamma\alpha\beta}
  \ ,
\end{equation}
where $g^{\beta\gamma}$
is the inverse of the spatial metric $g_{\alpha\beta}$
(which in general is not the same as the spatial components
of the coordinate-frame spacetime metric $g_{\kappa\lambda}$).
The trace-free spatial part of
equations~(\ref{torsionconstraintfaltformt})
defines the 6 identities,
\begin{equation}
\label{idfaltformtS}
  \mbox{6 ids:}
  \quad
  S_{\gamma\alpha\beta}
  +
  g_{\gamma[\alpha} S^\delta_{\beta]\delta}
  =
  \kappa
  \bigl(
  \tilde{\Spin}_{\gamma\alpha\beta}
  +
  g_{\gamma[\alpha} \tilde{\Spin}^\delta_{\beta]\delta}
  \bigr)
  \ .
\end{equation}

\subsection{Gravitational magnetic field}
\label{gravmag-sec}

The 6 identities~(\ref{idfaltformtS}) can be interpreted
as defining 6 redundant components of the spatial Lorentz connection
$\bGamma_{\alphaform}$
in terms of spatial exterior derivatives
$\dext_{\alphaform} \be_{\alphaform}$
of the spatial line interval.
These 6 redundant components
can be called the gravitational magnetic field,
equation~(\ref{gravmag}),
since the situation is analogous to that in electromagnetism,
where the 3-component magnetic field is redundant because
it can be replaced by the spatial exterior derivative of the spatial
components of the electromagnetic potential,
equation~(\ref{constraintelecformA}).

Let
$\Gamma_{\kappa\lambda\mu}$
denote the 24 components of the Lorentz connections projected
into the coordinate frame,
\begin{equation}
\label{Gammacoord}
  \Gamma_{\kappa\lambda\mu}
  \equiv
  e^k{}_\kappa e^l{}_\lambda \Gamma_{kl\mu}
  \ .
\end{equation}
The coordinate-frame Lorentz connections
$\Gamma_{\kappa\lambda\mu}$
inherit antisymmetry in their first two indices $\kappa\lambda$
from the antisymmetry of the Lorentz connections
$\Gamma_{kl\gamma}$ in their first two indices $kl$.
The 24 coordinate-frame Lorentz connections
$\Gamma_{\kappa\lambda\mu}$
are {\em not\/} the same as the Christoffel symbols
(of which there are 64, or 40 if torsion vanishes).

In accordance with its definition~(\ref{SCartan}),
the spatial torsion $\bS_{\alphaform}$
in equation~(\ref{torsionconstraintfaltformtS})
involves
$\tfrac{1}{2} [ \bGamma , \be ]_{\alphaform}$,
which comprises the 12 coordinate-frame Lorentz connections
$\Gamma_{\kappa[\beta\gamma]}$
antisymmetrized over the last two spatial indices $\beta\gamma$,
\begin{equation}
  \tfrac{1}{2} [ \bGamma , \be ]_{\alphaform}
  =
  \bGamma_{\alphaform} \cdot \be_{\alphaform}
  =
  -2
  \Gamma_{\kappa[\beta\gamma]}
  \, \bgamma^\kappa
  \, \ddi{2} x^{\beta\gamma}
  \ ,
\end{equation}
the sum over $\beta\gamma$ being over distinct pairs of spatial indices.
The 9 all-spatial components
$\Gamma_{\alpha\beta\gamma}$
of the coordinate-frame Lorentz connections are invertibly related to
the 9 all-spatial components $\Gamma_{\alpha[\beta\gamma]}$ by
\begin{equation}
\label{Gammaa}
  \Gamma_{\alpha\beta\gamma}
  =
  \Gamma_{\alpha[\beta\gamma]}
  +
  \Gamma_{\beta[\gamma\alpha]}
  -
  \Gamma_{\gamma[\alpha\beta]}
  \ .
\end{equation}
The coordinate-frame spatial Lorentz connections
$\Gamma_{\alpha\beta\gamma}$
resolve into a 3-component trace and a 6-component trace-free part,
\begin{equation}
\label{GammaBpi}
  \Gamma_{\alpha\beta\gamma}
  \equiv
  g_{\gamma[\alpha}
  \ddual{\pi}_{\beta]}
  +
  B_{\alpha\beta\gamma}
  \ .
\end{equation}
The trace part is the BSSN \cite{Baumgarte:1998te} momentum variable,
\begin{equation}
\label{trGammaa}
  \Gamma_{\alpha\delta}{}^\delta
  \equiv
  g^{\beta\gamma}
  \Gamma_{\alpha\beta\gamma}
  =
  -
  \ddual{\pi}_\alpha
  \ ,
\end{equation}
where
$g^{\beta\gamma}$
is the inverse of the spatial metric $g_{\alpha\beta}$,
and $\ddual{\pi}_\alpha$ is the spatial double-dual
(Appendix~\ref{spatialdual-sec})
of the all-spatial coordinate-frame momentum
$\pi_{\alpha\beta\gamma \delta\epsilon} \, \bgamma^\alpha \wedgie \bgamma^\beta \wedgie \bgamma^\gamma \, \ddi{2} x^{\delta\epsilon}$.
The trace-free part of the coordinate-frame spatial Lorentz connection
$\Gamma_{\alpha\beta\gamma}$
defines the 6-component gravitational magnetic field
$B_{\alpha\beta\gamma}$,
\begin{equation}
\label{gravmag}
  B_{\alpha\beta\gamma}
  \equiv
  \Gamma_{\alpha\beta\gamma}
  +
  g_{\gamma[\alpha}
  \Gamma_{\beta]\delta}{}^\delta
  \ .
\end{equation}
The gravitational magnetic field
$B_{\alpha\beta\gamma}$, equation~(\ref{gravmag}),
can be written alternatively as the spatial dual
(Appendix~\ref{spatialdual-sec})
of a symmetric $3 \times 3$ spatial matrix
$B_{\delta\gamma}$,
\begin{equation}
\label{gravmagalt}
  B_{\alpha\beta\gamma}
  =
  \varepsilon_{\alpha\beta}{}^\delta
  B_{\delta\gamma}
  \ .
\end{equation}

The 12-component time component
$\bpi_{\tform}$
of the momentum can be expressed as a linear combination
of the 6-component gravitational field,
the 6-component gauge field $\bGamma_{\tform}$,
and the 12-component spatial momentum $\bpi_{\alpha}$,
with coefficients depending on the 16 components of the vierbein $\be$.
The calculation is somewhat intricate,
and is relegated to Appendix~\ref{pitform-sec}.

A practical numerical way to calculate $\bpi_{\tform}$,
adopted by \cite{Hamilton:2017qls},
is to use Singular Value Decomposition (SVD).
The 12 components of $\bpi_{\tform}$
are determined by 6 gauge conditions and
the 12 constraints and identities~(\ref{torsionconstraintfaltformtS}).
This is 18 equations for 12 unknowns,
including 6 constraints whose validity is in principle guaranteed by
conservation of angular momentum.
SVD finds a solution to an overcomplete set of linear
equations essentially by discarding redundant linear
combinations of the equations.

\subsection{Expansion}

After the 3+1 split,
equation~(\ref{dvolume})
is an equation of motion for the spatial 3-volume $\be^3_{\alphaform}$,
\begin{equation}
\label{dtvolume}
  \dext_{\tform} \be^3_{\alphaform}
  +
  \dext_{\alphaform} \be^3_{\tform}
  -
  \be_{\tform} \cdot \bvartheta_{\alphaform}
  -
  \be_{\alphaform} \cdot \bvartheta_{\tform}
  =
  \kappa
  ( \be^2 \wedgie \tilde{\bSpin} )_{\tform}
  \ .
\end{equation}

Equation~(\ref{dexpansionform})
is an equation of motion for the spatial expansion $\bvartheta_{\alphaform}$
\begin{align}
\label{dtexpansionform}
  &
  \dext_{\tform} \bvartheta_{\alphaform}
  +
  \dext_{\alphaform} \bvartheta_{\tform}
  +
  \tfrac{1}{4}
  \left( \be^2
  \wedgie
  [ \bGamma , \bGamma ]
  \right)_{\tform}
\\
\nonumber
  &=
  -
  {\kappa \over 2}
  \left(
  \be_{\alphaform} \wedgie \tilde{\bT}_{\tform}
  +
  \be_{\tform} \wedgie \tilde{\bT}_{\alphaform}
  +
  \tilde{\bSpin}_{\tform} \wedgie \bpi_{\alphaform}
  +
  \tilde{\bSpin}_{\alphaform} \wedgie \bpi_{\tform}
  \right)
  \ .
\end{align}
Whereas the equation of motion~(\ref{Einsteineqaltformt})
for the spatial momentum $\bpi_{\alphaform}$
is sourced only by the time component
$\tilde{\bT}_{\tform}$
of the energy-momentum pseudovector 3-form,
the equation of motion~(\ref{dtexpansionform})
for the spatial expansion $\bvartheta_{\alphaform}$
is sourced by
$\be_{\alphaform} \wedgie \tilde{\bT}_{\tform} + \be_{\tform} \wedgie \tilde{\bT}_{\alphaform}$,
which depends in addition on the spatial component $\tilde{\bT}_{\alphaform}$
of the energy-momentum.
Thus direct numerical integration of the equation of motion~(\ref{dtexpansionform})
for the spatial expansion $\bvartheta_{\alphaform}$ will yield a slightly different result from
integrating the spatial line interval $\be_{\alphaform}$ and spatial momentum $\bpi_{\alphaform}$
and then inferring the spatial expansion from its definition
$\bvartheta_{\alphaform} = \tfrac{1}{2} \be_{\alphaform} \wedgie \bpi_{\alphaform}$.

A modified version of the equation of motion~(\ref{dtexpansionform})
may be obtained by subtracting from the right hand side
an arbitrary factor $n/2$ times the wedge product
$\be_{\tform} \wedgie ( \bPi \,{-}\, \kappa \tilde{\bT} )_{\alphaform}$
of $\be_{\tform}$ with the Hamiltonian/momentum constraints
(it would also be possible to subtract
an arbitrary factor times
$( \bS - \kappa \tilde{\bSpin} )_{\alphaform} \wedgie \bpi_{\tform}$,
but that nicety will be glossed over here,
since usually spin angular-momentum and torsion
are taken to vanish identically),
\begin{align}
\label{dtexpansionformH}
  &
  \dext_{\tform} \bvartheta_{\alphaform}
  +
  \dext_{\alphaform} \bvartheta_{\tform}
  -
  \tfrac{1}{4}
  \left( \be^2
  \wedgie
  [ \bGamma , \bGamma ]
  \right)_{\tform}
  +
  \tfrac{1}{2} n \,
  \be_{\tform} \wedgie \bPi_{\alphaform}
\\
\nonumber
  &=
  -
  {\kappa \over 2}
  \left(
  \be_{\alphaform} \wedgie \tilde{\bT}_{\tform}
  +
  ( 1 - n )
  \be_{\tform} \wedgie \tilde{\bT}_{\alphaform}
  +
  \tilde{\bSpin}_{\tform} \wedgie \bpi_{\alphaform}
  +
  \tilde{\bSpin}_{\alphaform} \wedgie \bpi_{\tform}
  \right)
  \ .
\end{align}
If the arbitrary factor is $n = 0$,
the result recovers equation~(\ref{dtexpansionform}).
If the arbitrary factor is $n = 1$,
then the right hand side of equation~(\ref{dtexpansionformH})
depends only on the time component $\tilde{\bT}_{\tform}$
of the energy-momentum,
and the result of integrating the equation of motion for $\bvartheta_{\alphaform}$
will be equivalent to the result of integrating $\be_{\alphaform}$ and $\bpi_{\alphaform}$
and deducing $\bvartheta_{\alphaform}$ therefrom.
If the arbitrary factor is $n = 2$,
the result is the Raychaudhuri equation,
which is a key ingredient of singularity theorems.


\subsection{The 12 coordinates and 12 momenta}

A perhaps surprising aspect of the equations of motion~(\ref{eqaltformtRS})
is that there are $12+12$ of them
for the 12 gravitational coordinates $\be_{\alphaform}$
and their 12 conjugate momenta $\bpi_{\alphaform}$.
This is surprising because
one is familiar with the notion that the gravitational field has
6 degrees of freedom
(for example, the 2 scalar, 2 vector, and 2 tensor modes of cosmology;
although only the 2 tensor degrees of freedom are propagating).
In the ADM formalism,
the gravitational coordinates are the 6 components of the
spatial metric $g_{\alpha\beta}$,
and their 6 conjugate momenta
define the 6 components of the extrinsic curvature.
Where then do the extra 6 degrees of freedom in each of
$\be_{\alphaform}$ and $\bpi_{\alphaform}$
come from?
The answer is that the extra degrees of freedom are gauge degrees
of freedom associated with Lorentz transformations.

Recall that a motivating idea of this paper is that the environment
defines preferred Lorentz frames
(the most familiar example being the Cosmic Microwave Background).
The 12 components of the spatial line interval
$\be_{\alphaform} \equiv e_{k \alpha} \, \bgamma^k \, \dd x^{\alpha}$
encode not only the spatial coordinate metric
$g_{\alpha \beta} = e^k{}_{\alpha} e_{k \beta}$,
but also the boost and rotation of the vierbein with respect to the
preferred Lorentz frame.

\subsection{Conventional Hamiltonian}

After the 3+1 split,
the alternative Hilbert Lagrangian~(\ref{Lgaltform}) can be written
\begin{equation}
  L_\grav^\prime
  =
  {I \over \kappa}
  \oint_{t_{\rm i}}^{t_{\rm f}}
  \bpi_{\alphaform}
  \wedgie
  \be_{\tform}
  +
  {I \over \kappa}
  \int
  \bpi_{\alphaform}
  \wedgie
  \dext_{\tform} \be_{\alphaform}
  -
  H_\grav^\prime
  \ ,
\end{equation}
with conventional (not-super) Hamiltonian
\begin{equation}
\label{conventionalHamg}
  H_\grav^\prime
  =
  {I \over \kappa}
  \left(
  - \,
  \bpi_{\tform}
  \wedgie
  \bS_{\alphaform}
  +
  \bPi_{\alphaform}
  \wedgie
  \be_{\tform}
  \right)
  \ .
\end{equation}
The conventional Hamiltonian~(\ref{conventionalHamg})
is a sum of the gauge and redundant components
$\be_{\tform}$ and $\bpi_{\tform}$
of the coordinates and momenta
wedged with the components
$\bPi_{\alphaform}$ and $\bS_{\alphaform}$
of the curvature and torsion
that are subject to constraints and identities.

\subsection{Generalization to other dimensions}

The treatment in this section generalizes without obstacle
to any number $N \geq 3$ of spacetime dimensions.
The momentum $\bpi$ conjugate to the line interval $\be$
in $N$ spacetime dimensions is,
generalizing equation~(\ref{piformdef}),
the pseudovector $(N{-}2)$-form
\begin{equation}
\label{piNformdef}
  \bpi
  \equiv
  (-)^{N-3}
  \be^{N-3} \wedgie \bGamma
  \ .
\end{equation}
The curvature pseudovector $(N{-}1)$-form $\bPi$ is,
generalizing equations~(\ref{Piformdef}),
\begin{align}
\label{PiNformdef}
  \bPi
  &\equiv
  \be^{N-3} \wedgie \bR
  -
  \be^{N-4} \wedgie
  \bS \wedgie \bGamma
\nonumber
\\
  &=
  \dext \bpi
  +
  \tfrac{1}{2} [ \bGamma , \bpi ]
  -
  \tfrac{1}{4}
  \be^{N-3} \wedgie
  [ \bGamma , \bGamma ]
  \ .
\end{align}
There are $\tfrac{1}{2} N^2 ( N {+} 1 )$ Hamilton's equations,
comprising
$2 N ( N {-} 1 )$ equations of motion,
$\tfrac{1}{2} N ( N {-} 1 )$ Gaussian constraints,
$N$ Hamiltonian/momentum constraints,
and
$\tfrac{1}{2} N ( N {-} 1 ) ( N {-} 3 )$ identities.

\section{Strong hyperbolicity and the WEBB formalism}
\label{hyperbolicity-sec}

The ensemble of equations~(\ref{eqaltformd})
constitute an Exterior Differential System (EDS).
For such a system there is an algorithm,
called Cartan's Test \cite{Cartan:2001,Ivey:2016},
to decide whether the system is integrable.
A system is integrable if,
given an appropriate set of initial conditions,
a solution exists and is unique.
Appendix~\ref{cartantest-sec}
shows that the system~(\ref{eqaltformd}) passes Cartan's Test,
and is therefore integrable.

However,
integrability does not guarantee good numerical behavior,
if small errors in the initial conditions blow up exponentially.
A condition that guarantees good numerical behavior is
that the system be strongly hyperbolic
\cite{Kreiss:2001cu,Nagy:2004td,Hilditch:2013sba}.
Loosely speaking,
strong hyperbolicity requires that perturbations to initial conditions
propagate as waves $\sim \ee^{\im \omega t}$
rather than growing exponentially $\sim \ee^{\alpha t}$.

Strong hyperbolicity for a first-order system of partial differential equations
is defined as follows
\cite{Hilditch:2013sba}.
Let
$u_i$
denote a set of variables satisfying the first-order system
\begin{equation}
  {\partial u_i \over \partial t}
  +
  A^\alpha_{ij}
  {\partial u_j \over \partial x^\alpha}
  +
  \cdots
  =
  0
  \ ,
\end{equation}
where $\cdots$ does not involve derivatives of the variables.
The matrix $A^\alpha_{ij}$ for each spatial coordinate direction $\alpha$
is called the principal symbol of the system.
The system is called weakly hyperbolic if, for every direction $\alpha$,
all the eigenvalues of the principal symbol are real.
The system is called strongly hyperbolic if in addition, for every $\alpha$,
the eigenvectors of the principal symbol form a complete set,
and the eigenvector matrix and its inverse are uniformly bounded.
A key advantage of the BSSN formalism over the ADM formalism
is that BSSN is strongly hyperbolic whereas ADM is only weakly hyperbolic
\cite{Kreiss:2001cu,Nagy:2004td}.

The equations under consideration are
the 24 equations of motion~(\ref{eqaltformt})
for the 24 variables
$u_i = \{ e_{k\alpha} , \pi_{klm\alpha\beta} \}$
comprising the components of
$\be_{\alphaform}$ and $\bpi_{\alphaform}$.
The thing that complicates the analysis of the hyperbolicity of these
equations is that the 18-component spatial Lorentz connection
$\bGamma_{\alphaform}$
depends
not only on the 12-component spatial momentum
$\bpi_{\alphaform}$
but also on the 6-component gravitational magnetic field,
\S\ref{gravmag-sec},
which itself depends on spatial derivatives of the spatial coordinates
$\be_{\alphaform}$.
The term
$\dext_{\alphaform} \bpi_{\tform}$
in the Einstein equations~(\ref{Einsteineqaltformt})
then includes some second-order spatial derivatives of $\be_{\alphaform}$,
while the terms
$\tfrac{1}{2} [ \bGamma_{\alphaform} , \bpi_{\tform} ]$
and
$\tfrac{1}{4} \be_{\tform} \wedgie [ \bGamma_{\alphaform} , \bGamma_{\alphaform} ]$
are quadratic in first-order spatial derivatives of $\be_{\alphaform}$.

The difficulty can be overcome by promoting the gravitational magnetic field
$B_{\alpha\beta\gamma}$
to a set of 6 independent variables
governed by their own equation of motion.
The operation of promoting derivatives of variables to independent variables
and enlarging the system of differential equations is called prolongation.
The system obtained by prolonging the gravitational magnetic field
proves to be the WEBB formalism
\cite{Buchman:2005ub},
a system of tetrad-based equations proposed by
Buchman \& Bardeen
\cite{Buchman:2003sq}
based on the work of
Estabrook, Robinson \& Wahlquist
\cite{Estabrook:1996wa}.
\cite{Buchman:2003sq}
prove that the WEBB system is strongly hyperbolic.

\cite{Buchman:2003sq}
work with the all tetrad-frame Lorentz connections
$\Gamma_{klm} \equiv e_m{}^\mu \Gamma_{kl\mu}$,
which they decompose into various parts labeled with different letters.
The 9-component all-spatial part
$\Gamma_{abc}$
of the tetrad-frame Lorentz connection is labeled $N$.
{\em If\/} the ADM gauge condition~(\ref{admgauge}) is imposed,
so that the tetrad-frame spatial hyperplane coincides with the
coordinate-frame spatial hyperplane,
then spatial tetrad-frame vectors $A_a$ and
spatial coordinate-frame vectors $A_\alpha$
are related by the spatial vierbein and its inverse,
$A_\alpha = e^a{}_\alpha A_a$
and
$A_a = e_a{}^\alpha A_\alpha$.
The spatial tetrad-frame connection
$\Gamma_{abc}$
then decomposes
into a 3-component trace part
(which \cite{Buchman:2003sq} label $n$),
identified here as the BSSN variable
$\ddual{\pi}_b$,
and a 6-component trace-free part
identified here as the gravitational magnetic field
$B_{abc}$,
\begin{equation}
  \Gamma_{abc}
  =
  \varepsilon_{abd} N_{cd}
  =
  \eta_{c[a} \ddual{\pi}_{b]}
  +
  B_{abc}
  \ ,
\end{equation}
which is the tetrad-frame version of equation~(\ref{GammaBpi}).

The WEBB system is discussed further in \S\ref{webb-sec}.

\section{Comparison to ADM, BSSN, WEBB, and LQG}
\label{admbssnlqg-sec}

For comparison,
this section recasts some other formalisms into the language
of multivector-valued differential forms.

The Arnowitt-Deser-Misner (ADM) formalism
\cite{ADM:1959,ADM:1963},
provides the backbone for most modern implementations of numerical relativity.
The Baumgarte-Shapiro-Shibata-Nakamura (BSSN) formalism
founded on the work of
\cite{Baumgarte:1998te,Shibata:1995we}
has gained popularity
because it proves to have better numerical stability than ADM
when applied to problems such as the merger of two black holes
\cite{Baumgarte:2010}.
The Wahlquist-Estabrook-Buchman-Bardeen (WEBB)
\cite{Buchman:2005ub}
formalism is a system of tetrad-based equations proposed by
Buchman \& Bardeen
\cite{Buchman:2003sq}
based on the work of
Estabrook, Robinson \& Wahlquist
\cite{Estabrook:1996wa}.

\ntab

Table~\ref{ntab}
summarizes for each approach considered in this paper
the number of equations of motion
(equations governing the evolution of variables with time),
the number of constraints
(equations that must be satisfied in the initial conditions
but are thereafter guaranteed),
the number of identities
(equations defining redundant variables in terms of others),
and the total number of variables.

\subsection{ADM gauge condition}
\label{ADMgauge-sec}

ADM and BSSN start by supposing that spacetime is sliced
into spatial surfaces of constant time $t$.
The unit normal to the spatial surfaces defines the tetrad time axis
$\bgamma^0$ at each point of spacetime.
Usually ADM and BSSN work entirely in a coordinate frame,
without reference to tetrads;
but there is no loss of generality to introduce a locally inertial
tetrad $\bgamma^k$
and to project into a coordinate frame at a later stage.
The condition that the tetrad time axis be normal to spatial
surfaces of constant time,
$\bgamma_0 \cdot \be_{\alphaform} = 0$,
imposes the 3 gauge conditions
\begin{equation}
\label{admgauge}
  e_{0{\alpha}} = 0
  \ .
\end{equation}
Physically,
the ADM gauge condition~(\ref{admgauge})
is equivalent to imposing that
the 3-dimensional coordinate spatial hyperplane coincides with
the 3-dimensional tetrad spatial hyperplane
(more formally,
the hyperplane spanned by the 3 spatial coordinate 1-forms
$\dd x^\alpha$
is the same as the hyperplane
spanned by the 3 spatial tetrad directions $\bgamma^a$).
The ADM gauge condition~(\ref{admgauge}) uses up the 3 degrees of
freedom of Lorentz boosts.
The ADM gauge choice~(\ref{admgauge})
is also a basic ingredient of Loop Quantum Gravity,
\S\ref{lqg-sec}.

\subsection{Double 3+1 split}
\label{double3plus1split-sec}

To elucidate the structure of the ADM formalism,
it is convenient to extend the coordinate 3+1 split~(\ref{formsplit})
to a double 3+1 split of Lorentz as well as coordinate indices.
Thus a multivector form $\ba$
splits into 4 components
$\ba_{\zeroform\tform}$,
$\ba_{\zeroform\alphaform}$,
$\ba_{\aform\tform}$,
and
$\ba_{\aform\alphaform}$
that represent respectively the time-time, time-space, space-time,
and space-space components of the multivector form
(the $\zeroform\tform$, $\zeroform\alphaform$, $\aform\tform$, and $\aform\alphaform$
subscripts should be interpreted as labels, not indices),
\begin{equation}
\label{form0split}
  \ba =
  \ba_{\zeroform\tform} + \ba_{\zeroform\alphaform}
  +
  \ba_{\aform\tform} + \ba_{\aform\alphaform}
  \ .
\end{equation}
In this notation,
the ADM gauge condition~(\ref{admgauge}) is
\begin{equation}
\label{admgaugeform}
  \be_{\zeroform\alphaform}
  =
  0
  \ .
\end{equation}
The coordinate time components
$\be_{\zeroform\tform}$ and $\be_{\aform\tform}$
of the line interval are commonly called the lapse $\alpha$ and shift $\beta_a$,
\begin{equation}
\label{lapseshift}
  e_{0t}
  =
  - \alpha
  \ , \quad
  e_{at}
  =
  - \beta_a
  \ .
\end{equation}

The ADM gauge condition~(\ref{admgaugeform}) reduces
the number of degrees of freedom of the spatial line interval
$\be_{\alphaform}$
from 12 to $3 \times 3 = 9$,
and of the spatial area element
$\be^2_{\alphaform}$
from 18 to the same number, $3 \times 3 = 9$.
The 9 components of the spatial line interval and spatial area element
subject to the ADM gauge condition~(\ref{admgaugeform})
are invertibly related to each other.

Since the spatial line interval $\be_{\alphaform}$
and spatial area element $\be^2_{\alphaform}$
subject to the ADM gauge condition~(\ref{admgaugeform})
are invertibly related,
either $\be_{\aform\alphaform}$ or $\be^2_{\aform\alphaform}$
may be used as gravitational coordinates.
The momenta conjugate to the spatial area element
$\be^2_{\aform\alphaform}$
are the
$3 \times 3 = 9$ components of the
Lorentz connections
$- \bGamma_{\zeroform\alphaform} \equiv \Gamma_{a0\alpha} \, \bgamma^0 \wedgie \bgamma^a \, \dd x^\alpha$
with one Lorentz index the tetrad time index $0$,
also called the extrinsic curvature.
In the literature the extrinsic curvature is commonly denoted
by the symbol $K$,
\begin{equation}
  K_{a\alpha}
  \equiv
  \Gamma_{a0\alpha}
  \ .
\end{equation}
The extrinsic curvature is a spatial tensor,
that is,
it is a tensor with respect to the 6-dimensional spatial
subgroup of the 10-dimensional Poincar\'e group
consisting of spatial coordinate transformations (at fixed $t$)
and spatial rotations of the spatial tetrad.
If torsion vanishes, then
the extrinsic curvature is symmetric
(it equals its transpose, equation~(\ref{symmetric})).
The momenta conjugate to the spatial line interval
$\be_{\aform\alphaform}$ are the
9 components of
$\bpi_{\zeroform\alphaform} \equiv - \be_{\aform\alphaform} \wedgie \bGamma_{\zeroform\alphaform}$.
In conventional notation,
the 9 components of the double dual
$\ddual{\bpi}_{\aform\tform}$
of the conjugate momenta
$\bpi_{\zeroform\alphaform}$
are invertibly related to the 9 components of the extrinsic curvature by
\begin{equation}
\label{ddualpi}
  \ddual{\pi}_{at\alpha}
  =
  K_{a\alpha} - e_{a\alpha} K
  \ ,
\end{equation}
where $K \equiv K^a_a$ is the trace of the extrinsic curvature.

To facilitate comparison to the approach proposed in the present paper,
choose
the 9 coordinates and 9 momenta in the ADM or BSSN formalisms
to be
$\be_{\aform\alphaform}$ and $\bpi_{\zeroform\alphaform}$
(one could equally well choose $\be^2_{\aform\alphaform}$
and $- \bGamma_{\zeroform\alphaform}$).
Variation of the combined gravitational and matter actions with
respect to variations
$\delta \bpi_{\zeroform\alphaform}$ and $\delta \be_{\aform\alphaform}$
yields $9 + 9 = 18$ equations of motion for the spatial line interval
$\be_{\aform\alphaform}$
and their conjugate momenta
$\bpi_{\zeroform\alphaform}$,
\begin{subequations}
\label{eqADMformtRS}
\begin{alignat}{2}
\label{torsioneADMformtS}
  \mbox{9 eqs of mot:}
  &\quad&
  \bS_{\aform\tform}
  &=
  \kappa \tilde{\bSpin}_{\aform\tform}
  \ ,
\\
\label{EinsteineqADMformtR}
  \mbox{9 eqs of mot:}
  &\quad&
  \bPi_{\zeroform\tform}
  &=
  \kappa \tilde{\bT}_{\zeroform\tform}
  \ .
\end{alignat}
\end{subequations}

Variation of the action with respect to
$\delta \be_{\zeroform\alphaform}$
yields 3 equations of motion,
\begin{equation}
\label{momentumADMformtR}
  \mbox{3 eqs of mot:}
  \quad
  \bPi_{\aform\tform}
  =
  \kappa \tilde{\bT}_{\aform\tform}
  \ .
\end{equation}
Note that the ADM gauge condition $\be_{\zeroform\alphaform} = 0$
is a gauge condition,
fixed after equations of motion are derived,
so it is correct to vary $\be_{\zeroform\alphaform}$ in the action,
leading to equation~(\ref{momentumADMformtR}).
Equation~(\ref{momentumADMformtR}) is an equation of motion in the sense
that it involves a time derivative $\dext_{\tform} \bpi_{\aform\alphaform}$;
but $\bpi_{\aform\alphaform}$ is not one of the momenta
$\bpi_{\zeroform\alphaform}$
conjugate to the line interval
$\be_{\aform\alphaform}$,
so equation~(\ref{momentumADMformtR}) has a different status
from the $9 + 9$ equations of motion~(\ref{eqADMformtRS}).
As described in \S\ref{ADM-sec} and \S\ref{BSSN-sec} below,
the treatment of equation~(\ref{momentumADMformtR}) is what distinguishes
the ADM and BSSN formalisms:
ADM treats it as a constraint equation,
while BSSN retains it as an equation of motion.

Variation of the combined gravitational and matter action with respect to
$\delta \bpi_{\aform\alphaform}$,
$\delta \bpi_{\aform\tform}$,
$\delta \bpi_{\zeroform\tform}$,
$\delta \be_{\aform\tform}$
and
$\delta \be_{\zeroform\tform}$
yields 19 equations
involving only spatial derivatives,
namely 9 identities, 6 Gaussian constraints,
and 4 Hamiltonian/momentum constraints,
\begin{subequations}
\label{constraintADMformt}
\begin{alignat}{2}
\label{identityADMformt}
  \mbox{3 ids:}
  &\quad&
  \bS_{\zeroform\tform}
  &=
  \kappa \tilde{\bSpin}_{\zeroform\tform}
  \ ,
\\
\label{Gaussianconstraint0ADMformt}
  \mbox{3 Gauss:}
  &\quad&
  \bS_{\zeroform\alphaform}
  &=
  \kappa \tilde{\bSpin}_{\zeroform\alphaform}
  \ ,
\\
\label{GaussianconstraintADMformt}
  \mbox{3 Gauss + 6 ids:}
  &\quad&
  \bS_{\aform\alphaform}
  &=
  \kappa \tilde{\bSpin}_{\aform\alphaform}
  \ ,
\\
\label{momentumconstraintADMformt}
  \mbox{3 mom:}
  &\quad&
  \bPi_{\zeroform\alphaform}
  &=
  \kappa \tilde{\bT}_{\zeroform\alphaform}
  \ ,
\\
\label{EinsteinconstraintADMformt}
  \mbox{1 Ham:}
  &\quad&
  \bPi_{\aform\alphaform}
  &=
  \kappa \tilde{\bT}_{\aform\alphaform}
  \ .
\end{alignat}
\end{subequations}
The 3 identities~(\ref{identityADMformt})
are not equations of motion
(they involve no time derivatives),
despite having a form label $\tform$.
Explicitly,
\begin{align}
  \mbox{3 ids:}
  \quad
  \dext_{\alphaform} \be_{\zeroform\tform}
  +
  \tfrac{1}{2} [ \bGamma_{\aform\alphaform} , \be_{\zeroform\tform} ]
  +
  \tfrac{1}{2} [ \bGamma_{\zeroform\tform} &, \be_{\aform\alphaform} ]
  +
  \tfrac{1}{2} [ \bGamma_{\zeroform\alphaform} , \be_{\tform\alphaform} ]
\nonumber
\\
  &=
  \kappa \tilde{\bSpin}_{\zeroform\tform}
  \ .
\end{align}
As earlier, equation~(\ref{GaussianconstraintfaltformtS}),
the 3 Gaussian constraints contained in
equations~(\ref{GaussianconstraintADMformt})
comprise the subset
obtained by wedging with the spatial line interval
$\be_{\aform\alphaform}$,
\begin{equation}
\label{GaussianconstraintADMformtS}
  \mbox{3 Gauss:}
  \quad
  ( \be \wedgie \bS )_{\aform\alphaform}
  =
  \kappa ( \be \wedgie \tilde{\bSpin} )_{\aform\alphaform}
  \ .
\end{equation}

The combined set of 40 Hamilton's
equations~(\ref{eqADMformtRS})--(\ref{constraintADMformt})
are identical to the earlier equations~(\ref{eqaltformtRS}),
(\ref{EinsteinconstraintfaltformtR}), and~(\ref{torsionconstraintfaltformtS}),
but grouped differently.
The different grouping results because ADM and BSSN treat
$\be_{\zeroform\alphaform}$
not as a dynamical variable,
but rather as a gauge variable set to zero.


\subsection{ADM formalism}
\label{ADM-sec}

The ADM formalism treats the 3 equations~(\ref{momentumADMformtR})
as constraint equations,
which is justified because they arise from variation of the action
with respect to the 3 components of $\be_{\zeroform\alphaform}$,
which ADM treats as gauge variables.
Actually,
ADM simply discards equations~(\ref{momentumADMformtR}) unceremoniously.
If torsion vanishes,
then Lorentz gauge symmetry enforces
that the Einstein tensor is symmetric,
equation~(\ref{Tsym}),
in which case the 3 equations~(\ref{momentumADMformtR})
should equal the transpose of
the 3 momentum constraints~(\ref{momentumconstraintADMformt}),
and can be dropped.

Because ADM uses up the gauge freedom under Lorentz boosts
on the vierbein,
the gauge freedom can no longer be used on the connections
(the 3 components of $\bGamma_{\zeroform{\tform}}$
can no longer be treated as gauge variables).
Thus in ADM the 3 Gaussian constraints~(\ref{GaussianconstraintADMformtS})
cease to be constraints;
rather, they become identities.


Commonly,
after the 3 gauge freedoms under Lorentz boosts have been fixed
by the ADM gauge choice~(\ref{admgaugeform}),
the ADM equations are cast entirely in the coordinate frame.
This has the advantage that the remaining variables
are gauge-invariant under the remaining Lorentz symmetries,
namely spatial rotations.
After a 3+1 split,
the gravitational coordinates are
the 6 spatial components $g_{\alpha\beta}$ of the metric,
and their conjugate momenta are, modulo a trace term,
the components
$\Gamma_{\alpha 0 \beta}$
of the extrinsic curvature $- \bGamma_{\zeroform\alphaform}$
projected into the coordinate frame
(specifically,
the conjugate momenta are
$\Gamma_{\alpha 0 \beta} \,{-}\, g_{\alpha\beta} \Gamma^\gamma_{0\gamma}$,
cf.\ equation~(\ref{ddualpi})).
For vanishing torsion, as ADM assumes,
the extrinsic curvature
$\Gamma_{\alpha 0 \beta}$
is symmetric and therefore also has 6 components.
Thus in ADM there are $6+6 = 12$ equations of motion.
The remaining gravitational degrees of freedom are embodied
in the 18 spatial components of the coordinate-frame connections
(the Christoffel symbols, not Lorentz connections).
The 18 spatial coordinate connections are determined by 18 identities
relating them to spatial derivatives of the spatial metric.

In ADM, the 6 Gaussian constraints cease to appear explicitly.
Where have they gone?
The answer is that they have disappeared into the 6-component antisymmetric
part of the Einstein equations, which vanishes identically
if torsion vanishes, as ADM assumes.

Altogether,
the ADM equations comprise $6 + 6 = 12$ equations of motion,
18 identities,
3 momentum constraints,
1 Hamiltonian constraint,
and 6 Gaussian constraints absorbed into the
antisymmetric part of the Einstein equations,
a total of 40 Hamilton's equations.

\subsection{BSSN formalism}
\label{BSSN-sec}

The BSSN formalism follows ADM for the most part,
except that it
retains the 3 equations~(\ref{momentumADMformtR})
as equations of motion.
As a result, BSSN retains
the 3 Gaussian constraints~(\ref{GaussianconstraintADMformtS})
as constraints.

In all,
the BSSN equations constitute
$6 + 6 + 3 = 15$ equations of motion,
15 identities,
3 momentum constraints,
1 Hamiltonian constraint,
3 Gaussian constraints,
and 3 Gaussian constraints absorbed into the
antisymmetric spatial part of the Einstein equations,
a total of 40 Hamilton's equations.

\subsection{WEBB formalism}
\label{webb-sec}

The WEBB formalism
\cite{Estabrook:1996wa,Buchman:2003sq}
is a tetrad system of equations for numerical relativity
similar in spirit to the approach of the present paper.
The WEBB formalism replaces the 6 identities~(\ref{idfaltformtS})
defining the gravitational magnetic field
$B_{\alpha\beta\gamma}$
in terms of spatial derivatives of the spatial coordinates $\be_{\alphaform}$
by equations of motion,
essentially by replacing some of the equations for the spatial torsion
$\bS_{\alphaform}$
by their exterior derivatives, equation~(\ref{dtorsionWEBBformt}),
a process called prolongation.
Promoting the gravitational magnetic field to a set of independent variables
governed by their own equation of motion has the virtue that the resulting
system is strongly hyperbolic,
as proved by \cite{Buchman:2003sq}.

Because the WEBB approach works entirely in the tetrad frame,
the WEBB system of equations does not translate directly
into the language of multivector-valued differential forms
used in the present paper.
However,
if the ADM gauge condition~(\ref{admgauge}) is imposed,
so that the 3-dimensional tetrad-frame and coordinate-frame
spatial tangent planes coincide, then there is a direct translation.
Subject to the ADM gauge condition,
the WEBB system has 30 equations of motion,
\begin{subequations}
\label{eqWEBBformtS}
\begin{alignat}{2}
\label{torsionaltformtwebb}
  \mbox{12 eqs of mot:}
  &\quad&
  \bS_{\tform}
  &=
  \kappa
  \tilde{\bSpin}_{\tform}
  \ ,
\\
\label{EinsteineqWEBBformt}
  \mbox{9 eqs of mot:}
  &\quad&
  \bPi_{\zeroform\tform}
  &=
  \kappa \tilde{\bT}_{\zeroform\tform}
  \ ,
\\
\label{dtorsionWEBBformt}
  \mbox{9 eqs of mot:}
  &\quad&
  ( \dext \bS )_{\aform\tform}
  &=
  \kappa
  ( \dext \tilde{\bSpin} )_{\aform\tform}
  \ .
\end{alignat}
\end{subequations}
Equations~(\ref{torsionaltformtwebb})
govern the evolution of the 12 components $\be_{\alphaform}$
of the line interval.
Equations~(\ref{EinsteineqWEBBformt})
are equivalent to equation~(39) of 
\cite{Buchman:2003sq},
and
govern the evolution of the 9 components of the extrinsic curvature
$\bGamma_{\zeroform\alphaform}$.
The 9 equations~(\ref{dtorsionWEBBformt})
are equivalent to equation~(40) of 
\cite{Buchman:2003sq},
and constitute
3 equations equivalent to the BSSN equations~(\ref{momentumADMformtR})
plus
6 equations of motion for the gravitational magnetic field,
as follows from the decomposition~(\ref{GammaBpi})
of the 9 all-spatial Lorentz connections
$\bGamma_{\aform\alphaform}$
into the (dual of the) 3 all-spatial components $\bpi_{\aform\alphaform}$
of the conjugate momentum
and
the 6-component gravitational magnetic field
$\bB_{\aform\alphaform}$.

The 9 equations of motion~(\ref{dtorsionWEBBformt})
must be supplemented by the 9 initial conditions
$\bS_{\aform\alphaform} = \kappa \tilde{\bSpin}_{\aform\alphaform}$,
which comprise 3 Gaussian constraints
and 6 equations relating the gravitational magnetic field to
spatial derivatives of the spatial coordinates $\be_{\alphaform}$.
These 6 equations are constraints
since they must be arranged to be satisfied on the initial spatial hypersurface,
but are thereafter guaranteed by
the Bianchi identity~(\ref{bianchiidentityformSR}).
I call them the $B$ constraints.
Altogether the WEBB system has
16 constraints:
\begin{subequations}
\label{eqWEBBformtS}
\begin{alignat}{2}
\label{Gaussianconstraint0WEBBformt}
  \mbox{3 Gauss:}
  &\quad&
  \bS_{\zeroform\alphaform}
  &=
  \kappa \tilde{\bSpin}_{\zeroform\alphaform}
  \ ,
\\
\label{GaussianconstraintWEBBformt}
  \mbox{3 Gauss + 6 $B$:}
  &\quad&
  \bS_{\aform\alphaform}
  &=
  \kappa \tilde{\bSpin}_{\aform\alphaform}
  \ ,
\\
\label{momentumconstraintWEBBformt}
  \mbox{3 mom:}
  &\quad&
  \bPi_{\zeroform\alphaform}
  &=
  \kappa \tilde{\bT}_{\zeroform\alphaform}
  \ ,
\\
\label{EinsteinconstraintWEBBformt}
  \mbox{1 Ham:}
  &\quad&
  \bPi_{\aform\alphaform}
  &=
  \kappa \tilde{\bT}_{\aform\alphaform}
  \ .
\end{alignat}
\end{subequations}
In effect, WEBB replaces the 6 identities~(\ref{idfaltformtS})
by 6 equations of motion and 6 constraints.

\subsection{Loop Quantum Gravity}
\label{lqg-sec}

Having come this far, it is useful to comment briefly on Loop Quantum Gravity
(LQG; e.g.\ \cite{Thiemann:2007zz,Rovelli:2007,Rovelli:2008,Bojowald2008}).
The language of multivector-valued forms is perfectly suited
to LQG.

Like ADM and BSSN,
LQG starts by making the ADM gauge choice~(\ref{admgaugeform}).
Unlike ADM and BSSN,
LQG treats the remaining 3 Lorentz degrees of freedom
under spatial rotations as having central importance.
As remarked in the third paragraph of \S\ref{double3plus1split-sec},
the 9-component spatial line interval $\be_{\aform\alphaform}$
and 9-component spatial area element $\be^2_{\aform\alphaform}$
subject to the ADM gauge condition~(\ref{admgaugeform})
are invertibly related,
so either $\be_{\aform\alphaform}$ or $\be^2_{\aform\alphaform}$
may be used as gravitational coordinates.
LQG adopts the spatial area element $\be^2_{\aform\alphaform}$
as the gravitational coordinates.
The momentum conjugate to the spatial area element
$\be^2_{\aform\alphaform}$
is the 9-component extrinsic curvature
$\bGamma_{\zeroform\alphaform}$.
LQG keeps all of the
$9 + 9$ equations of motion~(\ref{eqADMformtRS}),
including their antisymmetric parts.

Like ADM, LQG discards the 3 BSSN equations~(\ref{momentumADMformtR})
as constraint equations
enforced by the symmetry of the time-space components of the Einstein tensor,
for vanishing torsion.
LQG retains
the 4 Hamiltonian/momentum constraints~(\ref{momentumconstraintADMformt})
and~(\ref{EinsteinconstraintADMformt}),
and
the 3 Gaussian constraints~(\ref{GaussianconstraintADMformtS}).

In the seminal paper that subsequently led to LQG,
Ashtekar \cite{Ashtekar:1987gu}
noticed that bivectors in 4 spacetime dimensions (and only in 4 dimensions)
have a natural complex structure.
The complex structure allows
the 24-component bivector 1-form Lorentz connections $\bGamma$
to be split into distinct
12-component complex right- and left-handed chiral parts
\begin{equation}
  \bGamma_\pm
  \equiv
  ( 1 \pm \gamma_5 ) \bGamma
  \ ,
\end{equation}
where
$\gamma_5 \equiv - \im I$
is the chiral operator,
which is minus the product of
the quantum-mechanical imaginary $\im$
and the pseudoscalar $I$ of the geometric algebra,
equation~(\ref{pseudoscalar}).
The chiral Cartan's equations~(\ref{RCartan}) are
\begin{equation}
\label{RLQGCartan}
  \bR_\pm
  \equiv
  ( 1 \pm \gamma_5 ) \bR
  =
  \dext \bGamma_\pm + \tfrac{1}{4} [ \bGamma_\pm , \bGamma_\pm ]
  \ .
\end{equation}
That is,
the right(left)-handed Riemann curvature
depends only on
the right(left)-handed Lorentz connection.
Ashtekar pointed out that the imaginary part of the
(right-handed, without loss of generality)
chiral Hilbert Lagrangian,
\begin{equation}
\label{Lchiralform}
  L_+
  \equiv
  -
  {I \over \kappa} \,
  \be^2
  \wedgie
  \bR_+
  \ ,
\end{equation}
vanishes if torsion vanishes,
and that the resulting Hamilton's equations of motion are unchanged
from those of general relativity,
if torsion vanishes.

Ashtekar proposed as fundamental gravitational variables
the 9 components of the complex chiral
spatial Lorentz connection $\bGamma_{+ \aform\alphaform}$,
and the 9 components of the spatial area element $\be^2_{\aform\alphaform}$.
These $9 + 9$ variables are called Ashtekar variables,
\begin{equation}
  \bGamma_{+ \aform\alphaform}
  \ , \quad \be^2_{\aform\alphaform}
  \quad
  \mbox{Ashtekar variables}
  \ .
\end{equation}
The spatial line interval $\be_{\aform\alphaform}$
is the double dual of the area element
$\be^2_{\aform\alphaform}$ in 3 spatial dimensions,
\begin{equation}
  \be_{\aform\alphaform}
  =
  \ddual{\be^2}_{\aform\alphaform}
  \ ,
\end{equation}
a trick that works once again only in 4 spacetime dimensions,
hence 3 spatial dimensions.

A key point is that
keeping only 9 of the 18 spatial components $\bGamma_{\aform\alphaform}$
of the Lorentz connection
means that 9 components are discarded,
which allows the 9 identities contained
in equations~(\ref{identityADMformt})--(\ref{GaussianconstraintADMformt})
to be discarded.
Likewise 3 of the 6 time components $\bGamma_{\aform\tform}$
of the Lorentz connection are kept, and 3 are discarded,
allowing the 3 Gaussian constraints~(\ref{GaussianconstraintADMformtS})
to be discarded.
In all, the Ashtekar equations
comprise $9 + 9$ equations of motion,
zero identities,
4 Hamiltonian/momentum constraints, and
3 Gaussian constraints,
a total of 25 equations.
The missing 15 equations,
compared to the original set of 40 Hamilton's equations,
are
the 3 BSSN equations~(\ref{momentumADMformtR}),
which
for vanishing torsion are enforced
by the antisymmetry of the momentum components of the Einstein equation,
and the 9 identities and 3 Gaussian constraints determining
the 12 discarded components of the Lorentz connection of opposite chirality.

The resulting simplification of the gravitational
equations of motion (there being zero identities),
and the polynomial form of quantities
in the Ashtekar variables $\be^2_{\aform\alphaform}$
and $\bGamma_{+ \aform\alphaform}$,
led to hopes that canonical quantization could proceed.
Unfortunately the device of complexifying the Lorentz connection
proved too high a price, and that hope was not realized.

Subsequently it began to be realized,
starting with \cite{Jacobson:1987qk},
that if gravity is treated as a gauge theory of
spatial Lorentz transformations,
then a gauge rotation (a spatial rotation) by $2\pi$ around a loop
will lead to nontrivial quantum mechanical consequences.
Thus Loop Quantum Gravity was born.

Later versions of LQG are based on a real Lagrangian
\cite{Thiemann:1996ay}
\begin{equation}
\label{LLQGform}
  L_{\rm LQG}
  \equiv
  -
  {I \over \kappa} \,
  \be^2
  \wedgie
  \bR_\beta
  \ ,
\end{equation}
in which the imaginary $\im$ in equation~(\ref{RLQGCartan})
is replaced by a (usually) real parameter $\beta$
called the Barbero-Immirzi parameter
\cite{Barbero:1994ap,Immirzi:1996di},
\begin{equation}
  \bR_\beta
  \equiv
  \left( 1 + {I \over \beta} \right) \bR
  \ .
\end{equation}

\section{Summary}

This paper has presented a Hamiltonian approach to numerical
relativity
in which the full 10-dimensional Poincar\'e symmetry group
of general relativity is kept manifest.
The symmetry group is kept manifest
by using the powerful mathematical language of multivector-valued
differential forms.

The most immediate practical advantage of the approach is that it allows
a wider range of gauge choices than the traditional
ADM \cite{Lehner:2001}
or BSSN \cite{Baumgarte:2010}
approaches.
Specifically,
the proposed approach allows gauge choices to be imposed
not only on the line interval but also
on the Lorentz connections.
The broader range of gauge options could be advantageous
in some challenging problems
such as the interior structure of accreting, rotating black holes,
addressed in a subsequent paper
\cite{Hamilton:2017qls}.

In the proposed approach,
the 40 Hamilton's equations divide into
24 equations of motion and 6 identities,
along with 10 constraint equations.
The 6 identities define a 6-component gravitational magnetic field
analogous to the magnetic field of electromagnetism.
If the gravitational magnetic field is promoted to a set of independent
variables governed by their own equation of motion,
then the system becomes the WEBB
\cite{Estabrook:1996wa,Buchman:2003sq,Buchman:2005ub}
system with 30 equations of motion and 16 constraints,
which the authors of \cite{Buchman:2003sq}
have shown constitutes a strongly hyperbolic system.

It has been remarked that Loop Quantum Gravity
can be cast elegantly in the language of multivector-valued forms.

\begin{acknowledgements}
This research was supported in part by FQXI mini-grant FQXI-MGB-1626.
I thank Gavin Polhemus for the suggestion to use multivector-valued forms,
and Frank Estabrook for useful correspondence.
I thank Professor Fred Hehl for pointing out that the
$- \be \cdot \ddual{\bT}$
term in equation~(\ref{dSpinmform})
can be interpreted as the covariant exterior derivative of orbital angular momentum,
\S19(c) of \cite{Corson:1953}.
\end{acknowledgements}

\section*{References}

\bibliography{bh}

\appendix

\section{Notation}
\label{notation-sec}

\subsection{Indices}

Latin indices $k,l,...$ refer to Lorentz frames.
Greek indices $\kappa,\lambda,...$ refer to coordinate frames.
Early latin $a,b,...$
and greek $\alpha,\beta,...$
indices refer to spatial coordinates only.
Capital latin $K,L,...$ and greek $\Lambda,\Pi,...$
indices refer to distinct antisymmetric sequences of indices.
Implicit summation is over distinct antisymmetric sequences of indices,
since this removes ubiquitous factorial factors that would otherwise appear.

\subsection{Multivectors}

Multivectors
\cite{Doran:2003}
are elements of the Clifford algebra, or geometric algebra,
generated by basis elements (axes)
$\bgamma_k$
equipped with an inner product
\begin{equation}
  \bgamma_k \cdot \bgamma_l
  =
  \eta_{kl}
  \ ,
\end{equation}
and an antisymmetric outer product
$\bgamma_k \wedgie \bgamma_l = - \bgamma_l \wedgie \bgamma_k$.
In this paper there are 4 spacetime dimensions,
hence 4 axes
$\bgamma_k \equiv \{ \bgamma_0 , \bgamma_1 , \bgamma_2 , \bgamma_3 \}$,
and their inner product $\eta_{kl}$ is Minkowski.
The geometric algebra is identical to the Clifford algebra
of Dirac $\gamma$-matrices, hence the notation.

The metric generalizes to a multivector metric $\eta_{KL}$.
If the indices are
$K = k_1 ... k_n$
and
$L = l_1 ... l_n$,
then the multivector metric $\eta_{KL}$ is
\begin{equation}
  \eta_{KL}
  \equiv
  \bgamma_K \cdot \bgamma_L
  =
  (-)^{[n/2]} \eta_{k_1 l_1} \cdots \, \eta_{k_n l_n}
  \ ,
\end{equation}
which has an extra sign $(-)^{[n/2]}$.
Multivector indices are lowered and raised with the
multivector metric and its inverse.
A similar convention applies to forms:
there is an extra sign factor of $(-)^{[p/2]}$
when $p$ indices of a form are lowered or raised.

The pseudoscalar $I$ is the highest grade element of the geometric algebra,
\begin{equation}
\label{pseudoscalar}
  I
  \equiv
  \bgamma_0 \wedgie \bgamma_1 \wedgie \bgamma_2 \wedgie \bgamma_3
  \ .
\end{equation}
The square of the pseudoscalar in 4 spacetime dimensions is $-1$,
and in that sense behaves like an imaginary number,
\begin{equation}
  I^2 = -1
  \ .
\end{equation}

\subsection{Multiplication}

In this paper,
the product of 2 multivector forms $\ba$ and $\bb$
is the geometric product of the multivectors,
and the exterior product of the forms.
The wedge symbol is {\em not\/} used to denote the exterior product of forms,
since that notation would conflict with the use of the same symbol
for the outer product of multivectors.
Rather, all form products are by default exterior products.

The wedge symbol $\wedgie$ denotes the highest grade element
of the geometric product of multivector forms.
Thus the wedge product
$\ba \wedgie \bb$
of multivector forms $\ba$ and $\bb$ of grades $m$ and $n$
is the grade $m + n$ component of their geometric product.

The dot symbol $\cdot$ denotes the lowest grade element
of the geometric product of multivector forms
(except that the dot product of a scalar, a multivector of grade 0,
with a multivector is zero).
Thus the dot product
$\ba \cdot \bb$
of multivector forms $\ba$ and $\bb$ of nonzero grades $m$ and $n$
is the grade $| m - n |$ component of their geometric product.
In accordance with the convention of this paper,
$\ba \cdot \bb$
is a dot product of multivectors, but an exterior product of forms.

Commutators $[ \ba , \bb ]$ of multivector forms are defined in the usual way,
\begin{equation}
\label{commutatorab}
  [ \ba , \bb ]
  \equiv
  \ba \bb - \bb \ba
  \ .
\end{equation}
In accordance with the convention of this paper,
the products on the right hand side of equation~(\ref{commutatorab})
are geometric products of multivectors, and exterior products of forms.
The commutator of a bivector form with a multivector form of grade $n$
is a multivector form of the same grade $n$.

\subsection{Transpose of a multivector form}

The transpose $\ba^\transpose$ of a multivector form
$\ba$
of grade $n$ and form index $p$
is defined to be the multivector form,
of grade $p$ and form index $n$,
with multivector and form indices transposed by the vierbein,
\begin{align}
\label{formtranspose}
  \ba^\transpose
  &=
  \left( a_{kl .. \mu\nu ..} \, \bgamma^k \wedgie \bgamma^l \wedgie .. \  \ddi{p} x^{\mu\nu ..} \right)^\transpose
\nonumber
\\
  &=
  e^k{}_\kappa e^l{}_\lambda \cdot\cdot \, e_m{}^\mu e_n{}^\nu \cdot\cdot  \,
  a_{kl .. \mu\nu ..} \, \bgamma^m \wedgie \bgamma^n \wedgie \cdot\cdot \  \ddi{n} x^{\kappa\lambda ..}
\nonumber
\\
  &=
  a_{\kappa\lambda .. mn ..} \, \bgamma^m \wedgie \bgamma^n \wedgie \cdot\cdot \  \ddi{n} x^{\kappa\lambda ..}
  \ ,
\end{align}
implicitly summed over distinct antisymmetric sequences
$kl..$, $mn..$, ${\kappa\lambda ..}$, ${\mu\nu ..}$
of indices.
Transposing a multivector form twice leaves it unchanged,
\begin{equation}
  ( \ba^\transpose )^\transpose
  =
  \ba
  \ .
\end{equation}

The transpose of a symmetric tensor $\ba$,
one satisfying
$a_{k{\lambda}} \equiv a_{kl} e^l{}_{\lambda} = a_{lk} e^l{}_{\lambda} \equiv a_{{\lambda}k}$,
is itself,
\begin{equation}
\label{symmetric}
  \ba^\transpose
  =
  ( a_{k{\lambda}} \, \bgamma^k \, \dd x^{\lambda} )^\transpose
  =
  a_{{\lambda}k} \, \bgamma^k \, \dd x^{\lambda}
  =
  a_{k{\lambda}} \, \bgamma^k \, \dd x^{\lambda}
  =
  \ba
  \ .
\end{equation}
As a particular example,
the line interval is symmetric,
$\be^\transpose = \be$,
because the tetrad metric is symmetric.

\subsection{Hodge duals}

Hodge dual operations are defined for both multivectors and forms.
The Hodge dual of a multivector $\ba$ of grade $n$ is $I \ba$,
which is a multivector of grade $N {-} n$ in $N$ spacetime dimensions.
The form dual of a $p$-form
$\ba = \ba_{\Lambda} \, \ddi{p} x^{\Lambda}$
with multivector coefficients $\ba_{\Lambda}$
is the multivector $(N{-}p)$-form
$\hodge\ba$ given by
\begin{equation}
\label{formdual}
  \hodge\ba
  =
  \hodge\ba_{\Pi} \,
  \ddi{N{-}p} x^{\Pi}
  =
  \varepsilon_{\Pi}{}^{\Lambda} \ba_{\Lambda}
  \, \ddi{N{-}p} x^{\Pi}
  \ ,
\end{equation}
where $\varepsilon^{\Pi\Lambda}$ is the totally antisymmetric
pseudoscalar coordinate tensor,
and the implicit summation
is over distinct antisymmetric sequences ${\Lambda}$ and ${\Pi}$
of respectively $p$ and $N {-} p$
coordinate indices.
The antisymmetric tensor is normalized to
$\varepsilon^{0...(N{{-}}1)} = 1$
in the tetrad frame,
which ensures that the pseudoscalar satisfies
$I = \varepsilon^K \bgamma_K$
implicitly summed over the single distinct antisymmetric sequence
$K = 0...(N{-}1)$ of $N$ indices.
The definition of the form dual is arranged such that
the operations of taking the dual and the transpose commute,
so that
\begin{equation}
  \hodge\ba
  =
  ( I ( \ba^\transpose ) )^\transpose
  \ .
\end{equation}
The double dual of a multivector form $\ba$,
both a multivector dual and a form dual,
is denoted with a double-asterisk overscript
$\ddual{\phantom{a}}$,
\begin{equation}
\label{ddualdef}
  \ddual{\ba}
  \equiv
  I ( \hodge\ba )
  =
  \hodge( I \ba )
  \ .
\end{equation}
The double dual of a double dual is the identity,
\begin{equation}
  \ddual{\ddual{\ba}}
  =
  \ba
  \ .
\end{equation}
The double-dual transpose of a multivector form can be written
\begin{equation}
\label{ddualdef}
  \ddual{\ba}^\transpose
  =
  I ( I \ba )^\transpose
  \ .
\end{equation}

\subsection{Coordinate-frame spatial dual}
\label{spatialdual-sec}

Equations~(\ref{GammaBpi}),
(\ref{trGammaa}),
(\ref{gravmagalt}),
and (\ref{Gammatmu})
involve a coordinate-frame spatial dual.
The dual involves applying the totally antisymmetric coordinate-frame spatial
tensor
$\varepsilon_{\alpha\beta\gamma}$,
which is defined by
\begin{equation}
  \varepsilon_{\alpha\beta\gamma}
  \equiv
  \varepsilon_{t\alpha\beta\gamma}
  =
  e_{kt} e_{l\alpha} e_{m\beta} e_{n\gamma} \varepsilon^{klmn}
  =
  | e | [\alpha\beta\gamma]
  \ ,
\end{equation}
where $| e |$ is the determinant of the $4 \times 4$ spacetime vierbein matrix.
To ensure that the dual of a dual is the identity,
the indices on $\varepsilon_{\alpha\beta\gamma}$ must be
lowered and raised with the spatial coordinate metric
$g_{\alpha\beta} \equiv e_{k\alpha} e_{l\beta} \eta^{kl}$
and its inverse
$g^{\alpha\beta}$
(which is not the same as the spatial components of the coordinate-frame
spacetime metric $g^{\kappa\lambda}$).
The inverse totally antisymmetric coordinate-frame spatial tensor
$\varepsilon^{\alpha\beta\gamma}$ is
\begin{equation}
  \varepsilon^{\alpha\beta\gamma}
  =
  | g |^{-1} | e | [\alpha\beta\gamma]
  \ ,
\end{equation}
where $| g |$
is the determinant of the $3 \times 3$ spatial metric matrix.

\section{Bianchi identities and conservation laws}
\label{bianchi-sec}

\subsection{Bianchi identities}

The Jacobi identity applied to the triple product of the
(coordinate and Lorentz) covariant derivative
implies the Bianchi identities
\begin{subequations}
\label{bianchiidentityform}
\begin{align}
\label{bianchiidentityformSR}
  \dext \bS
  +
  \tfrac{1}{2}
  [ \bGamma , \bS ]
  +
  \tfrac{1}{2}
  [ \be , \bR ]
  =
  0
  &
  \ ,
\\
\label{bianchiidentityformRR}
  \dext \bR
  +
  \tfrac{1}{2}
  [ \bGamma , \bR ]
  =
  0
  &
  \ .
\end{align}
\end{subequations}
The $\tfrac{1}{2} [ \be , \bR ]$ term
in the torsion Bianchi identity~(\ref{bianchiidentityformSR})
is a vector 3-form whose 16 components
comprise the antisymmetric part
of the 36-component Riemann tensor,
\begin{equation}
\label{eR}
  \tfrac{1}{2}
  [ \be , \bR ]
  =
  \be \cdot \bR
  =
  R_{[{\kappa\lambda\mu}]n}
  \, \bgamma^n \, \ddi{3} x^{\kappa\lambda\mu}
  \ .
\end{equation}
The Riemann tensor $\bR$ is itself determined
by its definition~(\ref{RCartan})
in terms of the Lorentz connection $\bGamma$
and its exterior derivative $\dext \bGamma$.

The contracted Bianchi identities,
obtained by wedging the Bianchi identities~(\ref{bianchiidentityform})
with the line interval $\be$,
are
\begin{subequations}
\label{contractedbianchiidentityformSRR}
\begin{align}
\label{contractedbianchiidentityformSR}
  \dext ( \be \wedgie \bS )
  +
  \tfrac{1}{2} [ \bGamma , \be \wedgie \bS ]
  -
  \tfrac{1}{2} [ \be^2 , \bR ]
  &=
  0
  \ ,
\\
\label{contractedbianchiidentityformRR}
  \dext ( \be \wedgie \bR )
  +
  \tfrac{1}{2} [ \bGamma , \be \wedgie \bR ]
  -
  \bS \wedgie \bR
  &=
  0
  \ .
\end{align}
\end{subequations}
The term $\tfrac{1}{2} [ \be^2 , \bR ]$ in the contracted torsion
Bianchi identity~(\ref{contractedbianchiidentityformSR}) is
a bivector 4-form whose 6 components comprise the antisymmetric
part of the Ricci tensor,
\begin{align}
\label{antisymmetricRicci}
  - \tfrac{1}{2} [ \be^2 , \bR ]
  &=
  \be \wedgie ( \be \cdot \bR )
  =
  \be \cdot ( \be \wedgie \bR )
\nonumber
\\
  &=
  -
  e_{k{\kappa}}
  e_{l{\lambda}}
  R_{[{\mu\nu}]}
  \, \bgamma^k \wedgie \bgamma^l \, \ddi{4} x^{\kappa\lambda\mu\nu}
  \ .
\end{align}

The contracted Bianchi identities~(\ref{contractedbianchiidentityformSR})
and~(\ref{contractedbianchiidentityformRR})
enforce conservation respectively of total angular momentum
and of total energy-momentum.

Combining the contracted Bianchi identity~(\ref{contractedbianchiidentityformRR})
with the torsion Bianchi identity~(\ref{bianchiidentityformSR})
yields the pseudovector 4-form identity for the curvature $\bPi$
defined by equation~(\ref{Piformdef}),
\begin{equation}
\label{contractedbianchiidentityformPiR}
  \dext \bPi
  +
  \tfrac{1}{2}
  [ \bGamma , \bPi ]
  -
  \tfrac{1}{2}
  [ \be , \bR ] \wedgie \bGamma
  +
  \tfrac{1}{4}
  \bS \wedgie [ \bGamma , \bGamma ]
  =
  0
  \ .
\end{equation}

\subsection{Conservation of angular momentum}
\label{conservationangularmomentum-sec}

Invariance of the action of a matter species under Lorentz transformations
implies the law~(\ref{dSpinmform}) of conservation of the
angular momentum of that matter species.
If the spin angular-momentum of a matter species vanishes,
then the conservation law implies that the energy-momentum tensor of
that matter species is symmetric, equation~(\ref{Tsym}).
The energy-momentum tensor is generically not symmetric for
a matter species with non-vanishing spin angular-momentum.
The actions of gauge fields such as electromagnetism
are independent of the Lorentz connection,
so gauge fields carry no spin angular-momentum.
Fermionic fields on the other hand do depend on the Lorentz connection,
and carry spin angular-momentum.

Under an infinitesimal Lorentz transformation
generated by the infinitesimal bivector
$\bepsilon = \epsilon_{kl} \, \bgamma^k \wedgie \bgamma^l$,
any multivector form $\ba$ whose multivector components transform like a tensor
varies as
\begin{equation}
\label{daLorentz}
  \delta \ba = \tfrac{1}{2} [ \bepsilon , \ba ]
  \ .
\end{equation}
In particular,
since the vierbein $e_{k{\kappa}}$ is a tetrad vector,
the variation of the line interval
$\be$
under an infinitesimal Lorentz transformation $\bepsilon$ is
\begin{equation}
\label{deLorentz}
  \delta \be = \tfrac{1}{2} [ \bepsilon , \be ]
  \ .
\end{equation}
The components of the Lorentz connection
$\bGamma$
do not constitute a tetrad tensor,
so its variation under a Lorentz transformation is not given by
equation~(\ref{daLorentz}).
Rather, the variation of the Lorentz connection follows from
a difference of tetrad-frame tensors,
\begin{equation}
\label{dGammaformcoordeq}
  \delta \DextL \ba - \DextL \delta \ba
  =
  \tfrac{1}{2} [ \delta \bGamma , \ba ]
  \ ,
\end{equation}
where $\DextL$
denotes the Lorentz-covariant exterior derivative
\begin{equation}
\label{DextL}
  \DextL \ba \equiv \dext \ba + \tfrac{1}{2} [ \bGamma , \ba ]
  \ .
\end{equation}
The variation $\delta \bGamma$ of the Lorentz connection
under an infinitesimal Lorentz transformation generated by $\bepsilon$
is then
\begin{equation}
\label{dGammaLorentz}
  \delta \bGamma
  =
  -
  \left(
  \dext \bepsilon
  +
  \tfrac{1}{2} [ \bGamma , \bepsilon ]
  \right)
  \ .
\end{equation}
Inserting the variations~(\ref{deLorentz}) and~(\ref{dGammaLorentz})
of the line interval and Lorentz connection into the
variation~(\ref{dSmform}) of the matter action,
integrating by parts, and dropping the surface term,
yields the law of conservation of angular momentum,
\begin{equation}
\label{dSpinmform}
  \dext \ddual\bSpin
  +
  \tfrac{1}{2}
  [ \bGamma , \ddual\bSpin ]
  -
  \be \cdot
  \ddual{\bT}
  =
  0
  \ .
\end{equation}
If the conservation law~(\ref{dSpinmform}) is summed over all matter species,
and $\ddual{\bSpin}$ is eliminated using
the relation~(\ref{torsionmodified})
and the equation of motion~(\ref{torsioneqaltform}),
then the contracted torsion
Bianchi identity~(\ref{contractedbianchiidentityformSR})
is recovered.

If the spin angular-momentum of a matter species vanishes,
then the conservation law reduces to
\begin{equation}
\label{Tsym}
  \be \cdot
  \ddual{\bT}
  =
  0
  \ ,
\end{equation}
which translates into the statement that the energy-momentum tensor
of the matter species is symmetric, $\bT^\transpose = \bT$.

\subsection{Conservation of energy-momentum}
\label{conservationenergyomomentum-sec}

Invariance of the action of a matter species under coordinate transformations
implies the law~(\ref{dTmformS}) of conservation of energy-momentum
of that matter species.

Any infinitesimal 1-form
$\bepsilon \equiv \epsilon_{\kappa} \, \dd x^{\kappa}$
generates an infinitesimal coordinate transformation
\begin{equation}
\label{dcoordform}
  x^{\kappa}
  \rightarrow
  x^{\kappa} + \epsilon^{\kappa}
  \ .
\end{equation}
The variation of any quantity $\ba$
with respect to an infinitesimal coordinate transformation $\bepsilon$
is, by construction of the Lie derivative, minus its Lie derivative,
$\delta \ba = - {\cal L}_{\bepsilon} \ba$,
with respect to the vector $\epsilon^{\kappa}$.
The Lie derivative of a form $\ba$
with respect to a 1-form $\bepsilon$ is given by
``Cartan's magic formula,''
\begin{equation}
\label{Cartanmagic}
  {\cal L}_{\bepsilon} \ba
  =
  \bepsilon \fdot ( \dext \ba )
  +
  \dext ( \bepsilon \fdot \ba )
  \ ,
\end{equation}
where $\fdot$ denotes the form inner product
(the form dot $\fdot$ is written slightly larger than the multivector dot $\cdot$
to distinguish the two).
Cartan's magic formula~(\ref{Cartanmagic}) holds also for multivector-valued
forms, since multivectors are scalars under coordinate transformations.
However, the magic formula encounters the difficulty that
the Lie derivative of a multivector-valued form is not a tetrad tensor.
Consequently neither ${\cal L}_{\bepsilon} \be$
nor ${\cal L}_{\bepsilon} \bGamma$
is a tetrad tensor.
However, as pointed out at the beginning of \S{5.2.1} of \cite{Hehl:1994ue},
the Lagrangian is a Lorentz scalar,
so in varying the Lagrangian 4-form $L$,
the exterior derivative $\dext$ can be replaced by the
Lorentz-covariant exterior derivative $\DextL$
defined by equation~(\ref{DextL}),
\begin{equation}
\label{CartanmagicL}
  {\cal L}_{\bepsilon} L
  =
  \LieL{}_{\bepsilon} L
  \equiv
  \bepsilon \fdot ( \DextL L )
  +
  \DextL ( \bepsilon \fdot L )
  \ .
\end{equation}
The advantage of this replacement is that the Lorentz-covariant Lie derivatives
of the line interval and Lorentz connection are then (coordinate and tetrad)
tensors,
and the resulting law of conservation of energy-momentum is manifestly
tensorial, as it should be.
The Lorentz-covariant variation $\delta \be$ of the line interval
under an infinitesimal coordinate transformation~(\ref{dcoordform})
generated by $\bepsilon$ is
\begin{equation}
\label{dxformcoord}
  \delta \be
  =
  -
  \LieL{}_{\bepsilon} \be
  =
  - \,
  \dext ( \bepsilon \fdot \be )
  -
  \tfrac{1}{2} [ \bGamma , \bepsilon \fdot \be ]
  -
  \bepsilon \fdot \bS
  \ .
\end{equation}
The Lorentz connection $\bGamma$ is a coordinate tensor
but not a tetrad tensor,
so the formula for the Lorentz-covariant Lie derivative
does not apply directly to the Lorentz connection.
Rather, the Lorentz-covariant variation
$\delta \bGamma$
of the Lorentz connection
follows from equation~(\ref{dGammaformcoordeq}), implying
\begin{equation}
  \tfrac{1}{2} [ \delta \bGamma , \ba ]
  =
  - \,
  \LieL{}_{\bepsilon} \DextL \ba
  +
  \DextL \LieL{}_{\bepsilon} \ba
  \ ,
\end{equation}
which leads to
\begin{equation}
\label{dGammaformcoord}
  \delta \bGamma
  \equiv
  -
  \LieL{}_{\bepsilon} \bGamma
  =
  -
  \bepsilon \fdot \bR
  \ .
\end{equation}
Inserting the variations~(\ref{dxformcoord}) and~(\ref{dGammaformcoord})
of the line interval $\be$ and Lorentz connection $\bGamma$
into the variation~(\ref{dSmform}) of the matter action,
integrating by parts,
and dropping the surface term,
yields the law of conservation of energy-momentum,
\begin{equation}
\label{dTmformS}
  ( \bepsilon \cdot \be )
  \wedgie \bigl(
  \dext \ddual{\bT} + \tfrac{1}{2} [ \bGamma , \ddual{\bT} ]
  \bigr)
  -
  ( \bepsilon \fdot \bS ) \wedgie \ddual{\bT}
  -
  \ddual{\bSpin} \wedgie ( \bepsilon \fdot \bR )
  =
  0
  \ .
\end{equation}
I don't know a way to recast equation~(\ref{dTmformS})
in multivector-forms notation
with the arbitrary 1-form $\bepsilon$ factored out,
but in components
equation~(\ref{dTmformS}) reduces to
\begin{equation}
\label{dTm}
  \mathring{\DD}^m
  T_{{\kappa}m}
  +
  T^{mn}
  K_{mn{\kappa}}
  +
  \tfrac{1}{2}
  \Spin^{{\lambda}mn} \,
  R_{{\lambda\kappa}mn}
  = 0
  \ .
\end{equation}
Here $K_{mn\kappa}$
are the components of the contortion bivector 1-form $\bK$ defined by
$\bGamma \equiv \mathring{\bGamma} + \bK$,
where $\mathring{\bGamma}$
(with a $\mathring{~}$ overscript)
is the torsion-free (Levi-Civita) Lorentz connection.
If the energy-momentum conservation law~(\ref{dTmformS})
is summed over all matter components,
and the total
spin angular-momentum $\ddual{\bSpin}$
and
energy-momentum $\ddual{\bT}$
eliminated in favour of torsion $\bS$ and curvature $\bR$
using Hamilton's equations~(\ref{eqaltform}),
then the law of conservation of total energy-momentum
recovers the contracted
Bianchi identity~(\ref{contractedbianchiidentityformRR}).

If the spin angular-momentum of a matter species vanishes,
then the energy-momentum conservation law~(\ref{dTmformS})
for that species reduces to
\begin{equation}
  \dext \ddual{\bT} + \tfrac{1}{2} [ \mathring{\bGamma} , \ddual{\bT} ]
  =
  0
  \ ,
\end{equation}
with $\mathring{\bGamma}$ the torsion-free Lorentz connection.

\section{The time component of momentum}
\label{pitform-sec}

In the numerical scheme,
the equations of motion~(\ref{eqaltformt})
determine the evolution of the 12 spatial coordinates
$\be_{\alphaform}$
and 12 spatial momenta
$\bpi_{\alphaform}$.
The 4 time components $\be_{\tform}$ of the coordinates
are gauge variables, the lapse and shift.
The 12 time components $\bpi_{\tform}$ of the momenta can be computed
by first inferring the spatial Lorentz connection $\bGamma_{\alphaform}$,
as elaborated below,
and then deducing $\bpi_{\tform}$ from the definition~(\ref{piformdef}),
\begin{equation}
  \bpi_{\tform}
  =
  - \,
  \be_{\tform} \wedgie \bGamma_{\alphaform}
  -
  \be_{\alphaform} \wedgie \bGamma_{\tform}
  \ .
\end{equation}

The 18 components $\Gamma_{kl\alpha}$
of the spatial Lorentz connection $\bGamma_{\alphaform}$
can be computed from their coordinate-frame components
$\Gamma_{\kappa\lambda\alpha}$,
equation~(\ref{Gammacoord}).
The 18 coordinate frame components
$\Gamma_{\kappa\lambda\alpha}$
split into 9 all-spatial components
$\Gamma_{\beta\gamma\alpha}$
and 9 time-space components
$\Gamma_{t\beta\alpha}$.

The 9 all-spatial components
$\Gamma_{\alpha\beta\gamma}$
are given
by equation~(\ref{GammaBpi})
in terms of the spatial momentum
$\bpi_{\alphaform}$
and
the gravitational magnetic field
$B_{\alpha\beta\gamma}$.
The 6-component gravitational magnetic field
$B_{\alpha\beta\gamma}$
itself,
\S\ref{gravmag-sec},
can be calculated
either from spatial derivatives of the spatial coordinates $\be_{\alphaform}$,
or in the WEBB formalism from its equation of motion.

The 9 components
$\Gamma_{t\beta\gamma}$ follow from
the spatial momentum $\bpi_{\alphaform}$
and from the all-spatial components
$\Gamma_{\alpha\beta\gamma}$ given by equation~(\ref{GammaBpi}).
The definition~(\ref{piformdef}) of the trivector 2-form $\bpi$
implies that
\begin{align}
\label{pit}
  &
  \pi_{t\alpha\beta \gamma\delta}
  \, \bgamma^t \wedgie \bgamma^\alpha \wedgie \bgamma^\beta \,
  \ddi{2} x^{\gamma\delta}
\nonumber
\\
  &=
  -
  (
  2 g_{t\gamma} \Gamma_{\alpha\beta\delta}
  +
  4 g_{\beta\gamma} \Gamma_{t\alpha\delta}
  )
  \, \bgamma^t \wedgie \bgamma^\alpha \wedgie \bgamma^\beta \,
  \ddi{2} x^{\gamma\delta}
  \ ,
\end{align}
where $g_{\kappa\lambda}$ is the coordinate-frame spacetime metric.
The summation in equation~(\ref{pit}) is over distinct sets of pairs
$\alpha\beta$ and $\gamma\delta$ of spatial indices.
Define $\bmu_t$ to be the 9-component spatial bivector 2-form
\begin{equation}
\label{mut}
  \bmu_t
  \equiv
  (
  \pi_{t\alpha\beta \gamma\delta}
  +
  2 g_{t\gamma} \Gamma_{\alpha\beta\delta}
  )
  \, \bgamma^\alpha \wedgie \bgamma^\beta \,
  \ddi{2} x^{\gamma\delta}
  \ .
\end{equation}
Further, define
the 1-component spatial trivector 3-form
$\bnu_t$ be the contraction of $\bmu_t$ with the spatial metric
$\bg \equiv g_{\alpha\beta} \, \bgamma^\alpha \, \dd x^\beta$,
\begin{equation}
  \bnu_t
  \equiv
  \tfrac{1}{2}
  \bg \wedgie \bmu_t
  \ .
\end{equation}
Then equation~(\ref{pit}) inverts to
\begin{equation}
\label{Gammatmu}
  \Gamma_{t\alpha\beta}
  \, \bgamma^\alpha \, \dd x^\beta
  =
  - \,
  \ddual{\bmu}_t
  +
  \bg \wedgie \ddual{\bnu}_t
  \ ,
\end{equation}
analogously to the inversion formula~(\ref{Gammapi})
for $\bGamma$ in terms of $\bpi$ and its contraction $\bvartheta$.
The double duals on the right hand side of equation~(\ref{Gammatmu})
are to be interpreted as double duals with respect to
the coordinate spatial metric $g_{\alpha\beta}$ and
its inverse $g^{\alpha\beta}$
(see Appendix~\ref{spatialdual-sec}).

\section{Cartan's Test for integrability}
\label{cartantest-sec}

The equations of motion~(\ref{eqaltformtRS})
along with the constraints and identities~(\ref{EinsteinconstraintfaltformtR})
and~(\ref{torsionconstraintfaltformtS}),
form an Exterior Differential System (EDS).
For such a system there is an algorithm,
called Cartan's Test \cite{Cartan:2001,Ivey:2016},
to decide whether the system is integrable.
A system is integrable if,
given an appropriate set of initial conditions,
a solution exists and is unique.

Cartan's algorithm starts by collecting the 4 coordinates $x^\kappa$
and variables,
here the $16+24$ components $e_{k\kappa}$ and $\Gamma_{kl\kappa}$
of the line interval and Lorentz connection,
into a single 44-dimensional manifold,
\begin{equation}
\label{manifoldeds}
  \{ x^\kappa , e_{k\kappa} , \Gamma_{kl\kappa} \}
  \ .
\end{equation}
The tangent space of this manifold is spanned by the 44 differentials
\begin{equation}
\label{dmanifoldeds}
  \{ \dd x^\kappa , \dd e_{k\kappa} , \dd \Gamma_{kl\kappa} \}
  \ .
\end{equation}
For brevity and ease of algebraic manipulation, it is advantageous
to treat the variables and their differentials as multivectors
$\be_\kappa \equiv e_{k\kappa} \bgamma^k$
and
$\bGamma_\kappa \equiv \Gamma_{kl\kappa} \bgamma^k \wedgie \bgamma^l$,
and likewise
$\dd \be_\kappa \equiv \dd e_{k\kappa} \bgamma^k$
and
$\dd \bGamma_\kappa \equiv \dd \Gamma_{kl\kappa} \bgamma^k \wedgie \bgamma^l$.
Cartan interprets the 44 differentials~(\ref{dmanifoldeds})
as 1-forms satisfying an exterior calculus,
that is, the exterior derivatives of the 1-forms are zero,
and their exterior products are antisymmetric.
The equations of motion~(\ref{eqaltformd})
are recast as relations between the 1-forms,
\begin{subequations}
\label{eqaltformeds}
\begin{align}
\label{torsioneqaltformeds}
  \dd \be_\kappa \dd x^\kappa
  +
  \tfrac{1}{2} [ \bGamma_\kappa , \be_\lambda ]
  \ddi{2} x^{\kappa\lambda}
  &=
  \kappa \tilde{\bSpin}_{\kappa\lambda}
  \ddi{2} x^{\kappa\lambda}
  \ ,
\\
\label{Einsteineqaltformeds}
  \dd \bpi_{\kappa\lambda}
  \ddi{2} x^{\kappa\lambda}
  +
  \bigl(
  \tfrac{1}{2} [ \bGamma_\kappa , \bpi_{\lambda\mu} ]
  &-
  \tfrac{1}{4} \be_\kappa \wedgie [ \bGamma_\lambda , \bGamma_\mu ]
  \bigr)
  \ddi{3} x^{\kappa\lambda\mu}
\nonumber
\\
  &
  =
  \kappa \tilde{\bT}_{\kappa\lambda\mu}
  \ddi{3} x^{\kappa\lambda\mu}
  \ ,
\end{align}
\end{subequations}
with implicit summation over distinct antisymmetric sequences of indices.
Equation~(\ref{torsioneqaltformeds}) is a vector of 4 relations between 2-forms,
while equation~(\ref{Einsteineqaltformeds}) is a pseudovector
of 4 relations between 3-forms.
As usual in this paper,
all products of forms are to be understood as exterior products,
the wedge symbol $\wedgie$ being reserved to indicate an outer product of multivectors.
After the EDS equations~(\ref{eqaltformeds}) have been solved,
yielding solutions $\be_\kappa ( x )$ and $\bGamma_\kappa ( x )$
for the variables
$e_{k\kappa}$ and $\Gamma_{kl\kappa}$
as functions of the spacetime coordinates $x^\kappa$,
then the differentials of the variables can be replaced by an expansion
in partial derivatives,
and the EDS equations~(\ref{eqaltformeds}) then
reproduce the equations of motion~(\ref{eqaltformd}).
For example,
$\dd \be_\kappa \dd x^\kappa$
in equation~(\ref{torsioneqaltformeds})
becomes equal to the usual exterior derivative $\dext \be$,
\begin{equation}
  \dd \be_\lambda \dd x^\lambda
  =
  {\partial \be_\lambda \over \partial x^\kappa} \dd x^\kappa \dd x^\lambda
  =
  \left(
  {\partial \be_\lambda \over \partial x^\kappa}
  -
  {\partial \be_\kappa \over \partial x^\lambda}
  \right)
  \ddi{2} x^{\kappa\lambda}
  =
  \dext \be
  \ ,
\end{equation}
where the sum in the second expression is over both $\kappa$ and $\lambda$,
while the sum in the third expression is over distinct antisymmetric pairs
$\kappa\lambda$ of indices.

To proceed,
project equations~(\ref{torsioneqaltformeds})
and~(\ref{Einsteineqaltformeds})
respectively along arbitrary coordinate directions $\dd x^\kappa$
and arbitrary coordinate planes $\ddi{2} x^{\kappa\lambda}$,
yielding a ``Pfaffian system'' of $16+24 = 40$ 1-form equations,
\begin{subequations}
\label{eqalt1formeds}
\begin{align}
\label{torsioneqalt1formeds}
  \dd \be_\kappa
  +
  \tfrac{1}{2} [ \bGamma_{[\kappa} , \be_{\lambda]} ]
  \dd x^\lambda
  &=
  \kappa
  \tilde{\bSpin}_{[\kappa\lambda]}
  \dd x^\lambda
  \ ,
\\
\label{Einsteineqalt1formeds}
  \dd \bpi_{\kappa\lambda}
  +
  \bigl(
  \tfrac{1}{2} [ \bGamma_{[\kappa} , \bpi_{\lambda\mu]} ]
  &
  -
  \tfrac{1}{4} \be_{[\kappa} \wedgie [ \bGamma_\lambda , \bGamma_{\mu]} ]
  \bigr)
  \dd x^\mu
\nonumber
\\
  &=
  \kappa
  \tilde{\bT}_{[\kappa\lambda\mu]}
  \dd x^\mu
  \ .
\end{align}
\end{subequations}
Cartan's Test states that the system~(\ref{eqalt1formeds})
is integrable if it is ``involutive'' and
``torsion-free'' (this is a different meaning of torsion).
Involutive means that the number of independent equations (here 40)
equals the number of variables (here 40).
Torsion-free means that the exterior derivatives
of equations~(\ref{eqalt1formeds}) vanish on the equations of motion.
The exterior derivatives of equations~(\ref{torsioneqalt1formeds})
are the $16 +24 = 40$ 2-form equations
\begin{subequations}
\label{ddepi}
\begin{align}
\label{dde}
  \bigl(
  \tfrac{1}{2} [ \dd \bGamma_{[\kappa} , \be_{\lambda]} ]
  -
  \tfrac{1}{2} [ \bGamma_{[\kappa} , \dd \be_{\lambda]} ]
  \bigr)
  \dd x^\lambda
  &
  =
  \kappa \dext \tilde{\bSpin}_{[\kappa\lambda]}
  \dd x^\lambda
  \ ,
\\
\label{ddpi}
  \bigl(
  \tfrac{1}{2} [ \dd \bGamma_{[\kappa} , \bpi_{\lambda\mu]} ]
  -
  \tfrac{1}{2} [ \bGamma_{[\kappa} , \dd \bpi_{\lambda\mu]} ]
  &-
  \tfrac{1}{4} \dd \be_{[\kappa} \wedgie [ \bGamma_\lambda , \bGamma_{\mu]} ]
\nonumber
\\
  +
  \tfrac{1}{2} \dd \be_{[\kappa} \wedgie [ \bGamma_{[\lambda} , \dd \bGamma_{\mu]} ]
  \bigr)
  \dd x^\mu
  &
  =
  \kappa \dext \tilde{\bT}_{[\kappa\lambda\mu]}
  \dd x^\mu
  \ .
\end{align}
\end{subequations}
But equations~(\ref{ddepi}) are guaranteed by
the Bianchi identities~(\ref{bianchiidentityformSR})
and~(\ref{contractedbianchiidentityformPiR})
for $\bS$ and $\bPi$.
That is,
the Bianchi identities~(\ref{bianchiidentityformSR})
and~(\ref{contractedbianchiidentityformPiR})
express $\dext \bS$
and $\dext \bPi$
as linear combinations of
$\bS$, $\bPi$, and the antisymmetric part
$[ \be , \bR ]$
of the Riemann tensor $\bR$.
When the equations of motion
$\bS = \kappa \tilde{\bSpin}$
and
$\bPi = \kappa \tilde{\bT}$
are substituted into the Bianchi identities,
the conclusion is that
$\dext \bS = \kappa \dext \tilde{\bSpin}$
and
$\dext \bPi = \kappa \dext \tilde{\bT}$,
which are none other than equations~(\ref{ddepi}).
Note that
the Bianchi identity~(\ref{bianchiidentityformSR})
with $\bS \rightarrow \kappa \tilde{\bSpin}$
essentially defines the contribution to the antisymmetric Riemann tensor
$[ \be , \bR ]$
from any component of spin angular-momentum $\tilde{\bSpin}$.
The exterior derivative equations~(\ref{ddepi})
thus hold on the equations of motion,
and are therefore torsion-free in Cartan's sense.
Since the system~(\ref{eqalt1formeds}) is both involutive and torsion-free,
it is integrable.

The solution to the Cauchy problem comprises an ``integral submanifold,''
a submanifold of the 44-dimensional manifold~(\ref{manifoldeds})
of coordinates and variables
(below, the submanifold will be shown to be 24-dimensional).
For an integrable system, the integral submanifold passing through
any point of the parent manifold is unique.
The tangent space to the integral submanifold has the defining property
that all the 1-forms of the Pfaffian system~(\ref{eqalt1formeds}) vanish
in the tangent space.
Cartan shows that the Cauchy problem reduces to a local problem
in an infinitesimal neighborhood of a point on the integral submanifold.
Cartan solves the Cauchy problem by showing that there exist solutions
for the variables as finite-order polynomials in the coordinates $x^\kappa$
with origin shifted to the point in question.
Cartan's solution shows that conditions on the tangent space
to the integral submanifold at a point suffice to guarantee polynomial solutions
over an infinitesimal open neighborhood of the point.

Cartan's approach to constructing a solution of the EDS is iterative.
First choose a generic
(what generic means is addressed below)
coordinate direction, call it $\dd x^1$,
and project the equations of motion~(\ref{eqalt1formeds})
along that direction.
The torsion equation~(\ref{torsioneqalt1formeds})
projected along the 1-form $\dd x^1$
yields a vector of 4 1-forms,
\begin{equation}
\label{de1}
  \dd \be_1
  +
  \tfrac{1}{2} [ \bGamma_{[1} , \be_{\lambda]} ]
  \dd x^\lambda
  =
  \kappa \tilde{\bSpin}_{1\lambda}
  \dd x^\lambda
  \ .
\end{equation}
The 4 equations~(\ref{de1}) define a 4-dimensional subspace of
the tangent space to the putative integral submanifold.
The Einstein equation~(\ref{Einsteineqalt1formeds})
projected along the 1-form $\dd x^1$
yields nothing, since the equations involve antisymmetrized
pairs $\kappa\lambda$ of indices, and the pair $11$ is not antisymmetric.
The number 4 of equations added at the first iteration
is called the first Cartan character, $c_1 = 4$.

Next,
choose a second generic coordinate direction, call it $\dd x^2$,
and project the equations of motion~(\ref{eqalt1formeds})
along the directions $\dd x^1$ and $\dd x^2$.
The projection of the torsion equations~(\ref{torsioneqalt1formeds})
yields 4 more equations similar to~(\ref{de1}) but with index 2 in place of 1.
The projection of the Einstein equations~(\ref{Einsteineqalt1formeds})
yields a vector of 4 1-form equations
\begin{align}
\label{dpi12}
  \dd \bpi_{12}
  +
  \bigl(
  \tfrac{1}{2} [ \bGamma_{[1} , \bpi_{2\mu]} ]
  &
  -
  \tfrac{1}{4} \be_{[1} \wedgie [ \bGamma_2 , \bGamma_{\mu]} ]
  \bigr)
  \dd x^\mu
\nonumber
\\
  &=
  \kappa
  \tilde{\bT}_{[12\mu]}
  \dd x^\mu
  \ .
\end{align}
The second iteration thus adds 8 1-form equations
to the 4 1-form equations of the first iteration,
bringing the total number of 1-form directions lying in the tangent
space of the integral submanifold to $4 + 8 = 12$.
The number 8 of equations added at the second iteration
defines the second Cartan character, $c_2 = 8$.

Finally,
choose a third generic coordinate direction, call it $\dd x^3$,
and project the equations of motion~(\ref{eqalt1formeds})
along the three directions $\dd x^1$, $\dd x^2$, and $\dd x^3$.
The projection yields 4 more torsion equations similar to~(\ref{de1}),
and 8 more Einstein equations similar to~(\ref{dpi12}),
bringing the total number of 1-form directions lying in the tangent
space of the integral submanifold to $4 + 8 + 12 = 24$.
The number 12 of equations added at the third iteration
defines the third Cartan character, $c_3 = 12$.

The three iterations define 3 successive Cauchy problems.
The first problem is to solve the 4 equations at the first iteration
along a 2-dimensional coordinate submanifold
(42-dimensional submanifold of the full
44-dimensional manifold~(\ref{manifoldeds}))
where the last 2 coordinates $x^3$ and $x^4$
are fixed (so $\dd x^3 = \dd x^4 = 0$).
The 4 equations govern the evolution of the 4 components $\be_1$
along the direction $x^2$.
The solution involves 4 initial conditions,
the values of the components of $\be_1$
along an initial line of $x^1$ at fixed $x^2$.
The second problem is to solve the 12 equations at the second iteration
along a 3-dimensional coordinate submanifold where the last coordinate $x^4$
is fixed (so $\dd x^4 = 0$).
The solution involves 12 initial conditions,
the values of the components of $\be_1$, $\be_2$, and $\bpi_{12}$
along an initial surface of $x^1$ and $x^2$ at fixed $x^3$.
The third problem is to solve the 24 equations at the third iteration
along the 4-dimensional submanifold of coordinates.
The solution involves 24 initial conditions,
the values of $\be_\alpha$ and $\bpi_{\alpha\beta}$
with $\alpha$, $\beta$ running over 1, 2, 3,
along an initial 3-dimensional hypersurface of $x^1$, $x^2$, and $x^3$
at fixed $x^4$.
The solution at the third iteration fills out all 4 spacetime dimensions,
so no further iteration is needeed.
The tangent space of the integral submanifold at the third iteration
has $4 + 8 + 12 = 24$ dimensions,
so the integral submanifold has dimension 24.
The Cartan characters are
\begin{equation}
  \{ c_1 , c_2 , c_3 \}
  =
  \{ 4 , 8 , 12 \}
  \ ,
\end{equation}
summing to the dimension 24 of the integral submanifold.
The number of initial conditions is
$4 + 12 + 24 = 40$,
which is as it should be for a system of 40 equations governing
40 variables.
In terms of Cartan characters, the number of initial conditions is
$3 c_1 + 2 c_2 + c_3 = 40$.
The initial conditions comprise 4 functions of a single variable $x^1$
along a line at fixed $x^2$, $x^3$, $x^4$,
plus 12 functions of 2 variables $x^1$, $x^2$
along a surface at fixed $x^3$, $x^4$,
plus 24 functions of 3 variables $x^1$, $x^2$, $x^3$
along a hypersurface at fixed $x^4$.

The above construction required that the coordinate directions
$\dd x^1$, $\dd x^2$, and $\dd x^3$ be chosen ``generically.''
Not all coordinate 1-forms are generic.
For example, the 1-form $\dd \phi$ of the azimuthal coordinate
in spherical polar coordinates is ill-defined at the poles.
As another example, if the coordinate 1-form is chosen to be null,
then the vierbein matrix $e_{k\kappa}$ will have zero and infinite eigenvalues
(albeit finite determinant).
Physically, if the coordinate 1--form is null,
then the locally inertial frame defined by the tetrad $\bgamma^k$
is moving through the coordinates at the speed of light.
For example, this happens at the horizons of stationary black holes
with respect to stationary coordinates.
These are called coordinate singularities.
The solution is to choose generic coordinate directions that are not null.
Of course there are also genuine singularities,
such as those inside black holes,
where the locally inertial description of spacetime breaks down
irretrievably, and beyond which coordinates cannot be continued.

In numerical relativity, one wishes to choose a coordinate system that
is everywhere well-behaved,
the 4 coordinate 1-forms $\dd x^\kappa$ being nowhere null.
This means choosing a coordinate system with 3 spatial coordinates $x^\alpha$
and 1 time coordinate $t$,
such that the spatial 1-forms $\dd x^\alpha$ are everywhere spacelike,
and the time 1-form $\dd t$ is everywhere timelike.
A coordinate system is well-behaved over a region of spacetime
if and only if the $4 \times 4$ matrix of vierbein coefficients $e_{k\kappa}$
is positive definite
(all eigenvalues of the vierbein matrix are strictly positive)
over that region.

Thus in practice the fourth coordinate $x^4$ of Cartan's construction
is chosen to be a timelike coordinate $t$.
There are 24 initial conditions comprising
the values of the variables $\be_\alpha$ and $\bpi_{\alpha\beta}$
on the initial spatial hypersurface of constant time $t$.
But in addition there are $4 + 12 = 16$ differential conditions
(constraints and identities)
on the variables $\be_\alpha$ and $\bpi_{\alpha\beta}$
over the initial hypersurface, a total of 40 initial conditions.

\section{Electromagnetism}
\label{electromagnetism-sec}

The purpose of this Appendix is to show that
the structure of the equations of electromagnetism
is quite similar to the structure of the gravitational equations
derived in the text.

\subsection{Electromagnetic potential and field}

Electromagnetism is a $U(1)$ gauge theory.
The coordinates of electromagnetism are a 1-form
$\bA \equiv A_{\nu} \, \dd x^{\nu}$
which arises as the connection of
an electromagnetic gauge-covariant exterior derivative,
\begin{equation}
  \Dext_A
  \equiv
  \dext
  +
  \im e \bA
  \ ,
\end{equation}
which acts on fields $\varphi$ of dimensionless charge $e$.
Under an electromagnetic $U(1)$ gauge transformation of the field $\varphi$
by some scalar $\theta$,
\begin{equation}
  \varphi \rightarrow \ee^{- \im e \theta} \varphi
  \ ,
\end{equation}
the electromagnetic gauge-covariant derivative of $\varphi$
is required to transform as
\begin{equation}
  \Dext_A \varphi \rightarrow \ee^{- \im e \theta} \Dext_A \varphi
  \ .
\end{equation}
This requires that the electromagnetic potential $\bA$
vary under a gauge transformation as
\begin{equation}
\label{Agauge}
  \bA \rightarrow \bA + \dext \theta
  \ .
\end{equation}

The commutator of the electromagnetic gauge-covariant derivative
defines the electromagnetic field 2-form
$\bF \equiv F_{\mu\nu} \, \ddi{2} x^{\mu\nu}$
(implicit sum over distinct antisymmetric pairs $\mu\nu$),
\begin{equation}
\label{FdAform}
  \bF
  \equiv
  \dext \bA
  \ ,
\end{equation}
which is invariant under electromagnetic gauge transformations.
In components, equation~(\ref{FdAform}) is
\begin{equation}
  F_{\mu\nu}
  \, \ddi{2} x^{\mu\nu}
  =
  \left(
  {\partial A_\nu \over \partial x^\mu} - {\partial A_\mu \over \partial x^\nu}
  \right)
  \ddi{2} x^{\mu\nu}
  \ .
\end{equation}

\subsection{Electromagnetic action}

The electromagnetic Lagrangian 4-form,
expressed in terms of the electromagnetic coordinates $\bA$
and their derivatives, is
\begin{equation}
\label{Lelecform}
  L_\elec
  =
  \tfrac{1}{2}
  \hodge( \dext \bA ) \wedgie \dext \bA
  \ .
\end{equation}
The electromagnetic Lagrangian~(\ref{Lelecform}) is
invariant under electromagnetic gauge transformations,
as it must be.
The units here are Heaviside (SI units with $\varepsilon_0 = \mu_0 = 1$);
in Gaussian units the right hand side of equation~(\ref{Lelecform})
would have an extra factor of $1/(4\pi)$.
The momenta conjugate to the coordinates $\bA$ are
\begin{equation}
  {\delta L_\elec \over \delta \dext \bA}
  =
  \hodge{\dext \bA}
  =
  \hodge{\bF}
  \ ,
\end{equation}
which is the 2-form dual
$\hodge{\bF}$
of the electromagnetic field $\bF$.
Recast in super-Hamiltonian form,
the electromagnetic Lagrangian~(\ref{Lelecform}) is
\begin{equation}
  L_\elec
  =
  \hodge{\bF} \wedgie \dext \bA
  -
  H_\elec
  \ ,
\end{equation}
with coordinates $\bA$, momenta $\hodge{\bF}$,
and super-Hamiltonian
\begin{equation}
  H_\elec
  =
  \tfrac{1}{2}
  \hodge{\bF} \wedgie \bF
  \ .
\end{equation}

The variation
of the electromagnetic action
with respect to the coordinates $\bA$ and momenta $\hodge\bF$ is
\begin{equation}
\label{dSelecform1}
  \delta S_\elec
  =
  \int
  \hodge\bF \wedgie \dext \delta \bA
  +
  \delta \hodge\bF \wedgie \dext \bA
  -
  \delta \hodge\bF \wedgie \bF
  \ .
\end{equation}
Integrating the $\hodge\bF \wedgie \dext \delta \bA$ term
in equation~(\ref{dSelecform1}) by parts brings the variation
of the electromagnetic action to
\begin{equation}
\label{dSelecform}
  \delta S_\elec
  =
  \oint
  \hodge\bF \wedgie \delta \bA
  +
  \int
  - \,
  \dext \hodge\bF
  \wedgie
  \delta \bA
  +
  \delta \hodge\bF \wedgie
  (
  \dext \bA
  -
  \bF
  )
  \ .
\end{equation}

The variation of the matter action with respect to the
electromagnetic coordinates $\bA$ defines the 3-form dual $\hodge{\bj}$
of the matter electric current 1-form $\bj$,
\begin{equation}
\label{dSelecmatform}
  \delta S_\mat
  =
  \int
  \hodge\bj
  \wedgie
  \delta \bA
  \ .
\end{equation}

Requiring that the variation
$\delta ( S_\elec + S_\mat )$
of the combined electromagnetic and matter actions vanish
with respect to arbitrary variations
$\delta \bA$
and
$\delta \hodge\bF$
of the electromagnetic coordinates and momenta,
subject to the condition that $\delta \bA$ vanishes on the boundary,
yields Hamilton's equations,
\begin{subequations}
\label{Hamiltoneqelecform}
\begin{align}
\label{HamiltoneqelecformF}
  \dext \hodge\bF
  &=
  \hodge\bj
  \ ,
\\
\label{HamiltoneqelecformA}
  \dext \bA
  &=
  \bF
  \ .
\end{align}
\end{subequations}
The first Hamilton equation~(\ref{HamiltoneqelecformF})
is a 4-component 3-form
comprising Maxwell's source-full equations.
The second Hamilton equation~(\ref{HamiltoneqelecformA})
is a 6-component 2-form
that enforces the relation~(\ref{FdAform})
between the electromagnetic field $\bF$ and the electromagnetic potential $\bA$.

Taking the exterior derivative of
the first Hamilton equation~(\ref{HamiltoneqelecformF})
yields,
since $\dext^2 = 0$,
the electric current conservation law
\begin{equation}
\label{currentconservationform}
  \dext \hodge\bj
  =
  0
  \ .
\end{equation}
Taking the exterior derivative of
the second Hamilton equation~(\ref{HamiltoneqelecformA})
yields
\begin{equation}
  \dext \bF = 0
  \ ,
\end{equation}
which comprises Maxwell's source-free equations.

\subsection{3+1 split}
\label{splitelecform-sec}

Splitting the variation~(\ref{dSelecform})
of the combined electromagnetic and matter action
into time and space parts yields
\begin{align}
\label{dSelecqformt}
  \delta ( S_\elec &+ S_\mat )
  =
  \left[
  \oint
  \hodge\bF \wedgie \delta \bA
  \right]_{t_{\rm i}}^{t_{\rm f}}
\\
\nonumber
  &
  +
  \int_{t_{\rm i}}^{t_{\rm f}}
  - \,
  (
  \dext \hodge\bF
  -
  \hodge\bj
  )_{\tform}
  \wedgie
  \delta \bA_{\alphaform}
  -
  (
  \dext \hodge\bF
  -
  \hodge\bj
  )_{\alphaform}
  \wedgie
  \delta \bA_{\tform}
\\
\nonumber
  &\quad\quad\quad
  + \,
  \delta \hodge\bF_{\alphaform} \wedgie
  (
  \dext \bA
  -
  \bF
  )_{\tform}
  +
  \delta \hodge\bF_{\tform} \wedgie
  (
  \dext \bA
  -
  \bF
  )_{\alphaform}
  \ .
\end{align}
After the 3+1 split,
the electromagnetic coordinates are the 3 spatial components $\bA_{\alphaform}$
of the electromagnetic potential 1-form,
and their conjugate momenta are the 3 spatial components
$\hodge{\bF}_{\alphaform}$
of the dual electromagnetic field,
which 3 components comprise the electric field.
Variation of the action with respect to the variations $\delta \bA_{\alphaform}$
and $\delta \hodge{\bF}_{\alphaform}$
of the 3 coordinates and their 3 conjugate momenta
yields $3 + 3$ equations of motion
that give time derivatives of the coordinates and momenta,
\begin{subequations}
\label{Hamiltoneqmotelecform}
\begin{alignat}{2}
\label{HamiltoneqmotelecformF}
  \mbox{3 eqs of mot:}
&\quad&
  ( \dext \hodge\bF )_{\tform}
  \equiv
  \dext_{\tform} \hodge\bF_{\alphaform}
  +
  \dext_{\alphaform} \hodge\bF_{\tform}
  &=
  \hodge\bj_{\tform}
  \ ,
\\
\label{HamiltoneqmotelecformA}
  \mbox{3 eqs of mot:}
&\quad&
  ( \dext \bA )_{\tform}
  \equiv
  \dext_{\tform} \bA_{\alphaform}
  +
  \dext_{\alphaform} \bA_{\tform}
  &=
  \bF_{\tform}
  \ .
\end{alignat}
\end{subequations}
Note that $\hodge{\bF}_{\alphaform}$ is dual to $\bF_{\tform}$,
and those components comprise the electric field;
while $\bF_{\alphaform}$ is dual to $\hodge\bF_{\tform}$,
and those components comprise the magnetic field.

The time component $\bA_{\tform}$ of the electromagnetic potential
can be regarded as a gauge field,
arbitrarily adjustable by an electromagnetic
gauge transformation~(\ref{Agauge}).
Variation of the action with respect to variations $\delta \bA_{\tform}$
of the gauge variable imply the constraint equation
\begin{equation}
\label{constraintelecformF}
  \mbox{1 constraint:}
  \quad
  ( \dext \hodge\bF )_{\alphaform}
  =
  \hodge\bj_{\alphaform}
  \ .
\end{equation}
The constraint equation~(\ref{constraintelecformF}) has the property that,
provided that it is satisfied in the initial conditions,
it is guaranteed thereafter by conservation of electric current.
Conservation of electric current is a consequence of
the $U(1)$ symmetry of electromagnetism.

Variation of the action with respect to the variation
$\delta \hodge\bF_{\tform}$ of the 3 components of the magnetic field
implies the 3 identities
\begin{equation}
\label{constraintelecformA}
  \mbox{3 ids:}
  \quad
  ( \dext \bA )_{\alphaform}
  =
  \bF_{\alphaform}
  \ .
\end{equation}
The identities~(\ref{constraintelecformA})
express the 3 components of the magnetic field $\bF_{\alphaform}$
as the spatial exterior derivative
of the spatial potential $\bA_{\alphaform}$.
Equation~(\ref{constraintelecformA}) is not an equation of motion
since it does not involve any time derivatives;
nor is it a constraint equation, since its continued satisfaction
is not enforced by any conservation law.
Rather, the identities~(\ref{constraintelecformA})
mean that the magnetic field $\bF_{\alphaform}$ is a redundant field
that can be eliminated by replacing it with $( \dext \bA )_{\alphaform}$.
The spatial potential $\bA_{\alphaform}$ is itself determined
by its equation of motion~(\ref{HamiltoneqmotelecformA}).
Although the magnetic field is redundant,
it is required for a covariant description of the 4-dimensional
2-form electromagnetic field.
Moreover,
the magnetic field
(or rather its dual $\hodge\bF_{\tform}$)
appears in the equation of motion~(\ref{HamiltoneqmotelecformF})
for the electric field.

In all,
the 10 Hamilton's equations of electromagnetism
comprise $3 + 3 = 6$ equations of motion,
1 constraint, and 3 identities.

In electromagnetism,
the Hamiltonian variables after a 3+1 split
are the 3-component spatial potential $\bA_{\alphaform}$
and its conjugate momentum the 3-component spatial electric field
$\hodge{\bF}_{\alphaform}$.
In place of the spatial potential $\bA_{\alphaform}$
it is possible
to work with the 3-component magnetic field
$\bF_{\alphaform} \equiv ( \dext \bA )_{\alphaform}$,
in which case the variables are the 6 components of the electric and magnetic fields
$\hodge{\bF}_{\alphaform}$ and $\bF_{\alphaform}$.
This is the traditional (century-old) approach.
The electric and magnetic fields are electromagnetic gauge-invariant,
but not coordinate gauge-invariant
(the electric and magnetic fields form a coordinate tensor,
not a set of coordinate scalars).
In general relativity,
the analogous situation is that
the Hamiltonian variables after a 3+1 split
are the 12-component spatial vierbein $\be_{\aform\alphaform}$
and its conjugate momentum the 12-component spatial momentum
$\bpi_{\aform\alphaform}$.
In place of these it is possible, if torsion vanishes,
to work with Lorentz gauge-invariant
variables, the 6-component spatial metric $g_{\alpha\beta}$,
its 6-component conjugate momentum the extrinsic curvature $K_{\alpha\beta}$,
and the 18-component spatial coordinate connections $\Gamma_{\alpha\beta\gamma}$
(Christoffel symbols).
This is the ADM formalism.
Thus the ADM formalism can be thought of as a general relativistic analog
of the traditional electromagnetic formalism.

\end{document}